\RequirePackage{ifpdf}
\ifpdf 
\documentclass[pdftex]{sigma}
\else
\documentclass{sigma}
\fi

\numberwithin{equation}{section}

\newcommand{\rd}{\partial}

\newcommand{\tp}[1]{\:{}^{\mathrm{t}}#1}

\newcommand{\ZZ}{\mathbf{Z}}
\newcommand{\bsa}{\boldsymbol{a}}
\newcommand{\bst}{\boldsymbol{t}}
\newcommand{\bsabar}{\bar{\bsa}}
\newcommand{\bstbar}{\bar{\bst}}
\newcommand{\bszero}{\boldsymbol{0}}
\newcommand{\abar}{\bar{a}}
\newcommand{\fbar}{\bar{f}}
\newcommand{\gbar}{\bar{g}}

\newcommand{\tbar}{\bar{t}}
\newcommand{\vbar}{\bar{v}}
\newcommand{\wbar}{\bar{w}}
\newcommand{\Abar}{\bar{A}}
\newcommand{\Bbar}{\bar{B}}
\newcommand{\Cbar}{\bar{C}}
\newcommand{\Dbar}{\bar{D}}
\newcommand{\Hbar}{\bar{H}}
\newcommand{\Kbar}{\bar{K}}
\newcommand{\Pbar}{\bar{P}}
\newcommand{\Sbar}{\bar{S}}
\newcommand{\Ubar}{\bar{U}}
\newcommand{\Vbar}{\bar{V}}
\newcommand{\Wbar}{\bar{W}}

\newcommand{\Psibar}{\bar{\Psi}}

\begin{document}

\allowdisplaybreaks

\renewcommand{\PaperNumber}{109}

\FirstPageHeading

\ShortArticleName{Pfaf\/f--Toda Hierarchy}

\ArticleName{Auxiliary Linear Problem, Dif\/ference Fay Identities\\
and Dispersionless Limit of Pfaf\/f--Toda Hierarchy}

\Author{Kanehisa TAKASAKI}

\AuthorNameForHeading{K. Takasaki}

\Address{Graduate School of Human and Environmental Studies,
Kyoto University,\\
Yoshida, Sakyo, Kyoto, 606-8501, Japan}

\Email{\href{mailto:takasaki@math.h.kyoto-u.ac.jp}{takasaki@math.h.kyoto-u.ac.jp}}

\URLaddress{\url{http://www.math.h.kyoto-u.ac.jp/~takasaki/}}

\ArticleDates{Received August 27, 2009, in f\/inal form December 15, 2009; Published online December 19, 2009}

\Abstract{Recently the study of Fay-type identities revealed some new features
of the DKP hierarchy (also known as ``the coupled KP hierarchy'' and
``the Pfaf\/f lattice'').  Those results are now extended to a Toda version
of the DKP hierarchy (tentatively called ``the Pfaf\/f--Toda hierarchy'').
Firstly, an auxiliary linear problem of this hierarchy is constructed.
Unlike the case of the DKP hierarchy, building blocks of the auxiliary
linear problem are dif\/ference operators.  A set of evolution equations
for dressing operators of the wave functions are also obtained.
Secondly, a system of Fay-like identities (dif\/ference Fay identities)
are derived.  They give a generating functional expression of
auxiliary linear equations.  Thirdly, these dif\/ference Fay identities
have well def\/ined dispersionless limit (dispersionless Hirota equations).
As in the case of the DKP hierarchy, an elliptic curve is hidden
in these dispersionless Hirota equations.  This curve is a kind of
spectral curve, whose def\/ining equation is identif\/ied with
the characteristic equation of a subset of all auxiliary linear equations.
The other auxiliary linear equations are related to quasi-classical
deformations of this elliptic spectral curve. }


\Classification{35Q58; 37K10}

\section{Introduction}

This paper is a sequel of the study on
Fay-type identities of integrable hierarchies,
in  particular the DKP hierarchy \cite{Takasaki07}.
The DKP hierarchy is a variant of the KP hierarchy
and obtained as a subsystem of Jimbo and Miwa's hierarchy
of the $D_\infty'$ type \cite{JM83,KvdL98}.
The same hierarchy was rediscovered later on
as ``the coupled KP hierarchy'' \cite{HO91}
and ``the Pfaf\/f lattice'' \cite{AHvM99,ASvM02,AvM02},
and has been studied from a variety of points of view
\cite{Kakei99,Kakei00,vdL01,AvM01,AKvM02,IWS02,IWS03,KM06,KP07}.
The term ``Pfaf\/f'' stems from the fact that Pfaf\/f\/ians
play a role in many aspects of this system.
The previous study \cite{Takasaki07} revealed
some new features of this relatively less known
integrable hierarchy.   In this paper,
we extend those results to a Toda version of the DKP hierarchy.

The integrable hierarchy in question is a slight modif\/ication
of the system proposed by Willox \cite{Willox02,Willox05}
as an extension of the Jimbo--Miwa $D_\infty'$ hierarchy.
We call this system, tentatively, ``the Pfaf\/f--Toda hierarchy''
(as an abbreviation of the ``Pfaf\/f\/ian'' or ``Pfaf\/f\/ianized''
Toda hierarchy).  Following the construction of Jimbo and Miwa,
Willox started from a fermionic def\/inition of the tau function,
and derived this hierarchy in a bilinear form.
The lowest level of this hierarchy contains
a $2+2D$ (2 continuous and 2 discrete) extension
\begin{gather*}
  \frac{1}{2}D_xD_y\tau(s,r,x,y)\cdot\tau(s,r,x,y)
  + \tau(s-1,r,x,y)\tau(s+1,r,x,y) \\
\qquad{} - \tau(s,r-1,x,y)\tau(s,r+1,x,y)
  = 0
\end{gather*}
of the usual $2+1D$ Toda equation
and an additional $2+2D$ equation
\begin{gather*}
  D_x\tau(s,r,x,y)\cdot\tau(s+1,r-1,x,y)
  + D_y\tau(s,r-1,x,y)\cdot\tau(s+1,r,x,y)
  = 0
\end{gather*}
(see the papers of Santini et al.~\cite{SND04},
Hu et al.~\cite{HLNY05} and Gilson and Nimmo \cite{GN05}
for some other sources of these equations).
Willox further presented an auxiliary linear problem
for these lowest equations, but extending it
to the full hierarchy was an open problem.
We f\/irst address this issue, then turn to issues
of Fay-like identities and dispersionless limit.

As we show in this paper, the Pfaf\/f--Toda hierarchy
is indeed a mixture of the DKP and Toda hierarchies.
Firstly, we can formulate an auxiliary linear problem
as a two-component system like that of the DKP hierarchy
\cite{Takasaki07}, but building blocks therein
are dif\/ference (rather than dif\/ferential) operators
as used for the Toda hierarchy.
Secondly, the dif\/ferential Fay identities
of the DKP hierarchy are replaced by
``dif\/ference Fay identities'' analogous to
those of the Toda hierarchy \cite{Zabrodin01,Teo06}.
Lastly, those dif\/ference Fay identities have
well def\/ined dispersionless limit to the so called
``dispersionless Hirota equations''.
These equations resemble the dispersionless
Hirota equations of the Toda hierarchy
\cite{Zabrodin01,KKMWWZ01,BMRWZ01,Teo03},
but exhibits a more complicated structure
parallel to the dispersionless Hirota equations
of the DKP hierarchy \cite{Takasaki07}.

Among these rich contents,
a particularly remarkable outcome
is the fact that an elliptic curve is hidden
in the dispersionless Hirota equations.
A similar elliptic curve was also encountered
in the dispersionless Hirota equations of
the DKP hierarchy \cite{Takasaki07},
but its true meaning remained to be clarif\/ied.
This puzzle was partly resolved by
Kodama and Pierce \cite{KP08}.  They interpreted
the curve as an analogue of the ``spectral curve''
of the dispersionless 1D Toda lattice.
We can now give a more def\/inite answer to this issue.
Namely, these curves are def\/ined by
the characteristic equations of a subset of
the full auxiliary linear equations,
hence may be literally interpreted as spectral curves.
The other auxiliary linear equations are related to
``quasi-classical deformations'' \cite{KKMA04,KKMA05}
of these curves.

This paper is organized as follows.
In Section~\ref{section2}, we formulate the Pfaf\/f--Toda hierarchy
as a~bilinear equation for the tau function.
This bilinear equation is actually a~generating
functional expression of an inf\/inite number of
Hirota equations.   In Section~\ref{section3}, we present
a full system of auxiliary linear equations
that contains Willox's auxiliary linear equations.
A~~system of evolution equations for ``dressing operators''
of the wave functions are also obtained.
The dressing operators are dif\/ference operators
in a direction ($s$-direction) of the 2D lattice;
another direction ($r$-direction)
plays the role of a discrete time variable.
Section~\ref{section4} deals with the dif\/ference Fay identities.
These Fay-like identities are derived
from the bilinear equation of Section~\ref{section2}
by specializing the values of free variables.
We show that they are auxiliary linear equations
in disguise, namely, they give a generating functional
expression of the auxiliary linear equations of Section~\ref{section3}.
Section \ref{section5} is devoted to the issues of dispersionless limit.
The dispersionless Hirota equations are derived
from the dif\/ferential Fay identities as a kind of
``quasi-classical limit''.  After rewriting
these dispsersionless Hirota equations,
we f\/ind an elliptic curve hidden therein,
and identify a set of auxiliary linear equations
for which the curve can be interpreted as
a spectral curve.

\section{Bilinear equations}\label{section2}

The Pfaf\/f--Toda hierarchy has two discrete variables
$s,r \in \ZZ$ and two sets of continuous variables
$\bst = (t_1,t_2,\ldots)$,
$\bstbar = (\tbar_1,\tbar_2,\ldots)$.
In this section, we present this hierarchy
in a bilinear form, which comprises
various bilinear equations for the tau function
$\tau = \tau(s,r,\bst,\bstbar)$.
In the following consideration,
we shall frequently use shortened notations
such as $\tau(s,r)$ for $\tau(s,r,\bst,\bstbar)$
to save spaces.

\subsection{Bilinear equation of contour integral type}\label{section2.1}

The most fundamental bilinear equation is the equation
\begin{gather}
 \oint\frac{dz}{2\pi i}
     z^{s'+r'-s-r}e^{\xi(\bst'-\bst,z)}
     \tau(s',r',\bst'-[z^{-1}],\bstbar')
     \tau(s,r,\bst+[z^{-1}],\bstbar) \nonumber\\
 \!\qquad\quad{} + \!
  \oint\frac{dz}{2\pi i}
     z^{s+r-s'-r'-4}e^{\xi(\bst-\bst',z)}
     \tau(s'+1,r'+1,\bst'+[z^{-1}],\bstbar')
     \tau(s-1,r-1,\bst-[z^{-1}],\bstbar) \nonumber\\
\qquad{}= \oint\frac{dz}{2\pi i}
     z^{s'-r'-s+r}e^{\xi(\bstbar'-\bstbar,z^{-1})}
     \tau(s'+1,r',\bst',\bstbar'-[z])
     \tau(s-1,r,\bst,\bstbar+[z])\nonumber\\
  \!\qquad\quad{} +
  \!\oint\frac{dz}{2\pi i}
     z^{s-r-s'+r'}e^{\xi(\bstbar-\bstbar',z^{-1})}
     \tau(s',r'+1,\bst',\bstbar'+[z])
     \tau(s,r-1,\bst,\bstbar-[z])
\label{bilin-tau}
\end{gather}
that is understood to hold for arbitrary values of
$(s,r,\bst,\bstbar)$ and $(s',r',\bst',\bstbar')$.
This equation is a~modif\/ication of the bilinear equation
derived by Willox \cite{Willox02,Willox05}
in a fermionic construction of the tau function
(see Section~\ref{section2.3} below).  Note that we have used
the standard notations
\begin{gather*}
   [z] = \left(z,\frac{z^2}{2},\frac{z^3}{3},\ldots\right), \qquad
   \xi(\bst,z) = \sum_{k=1}^\infty t_kz^k,
\end{gather*}
and both hand sides of the bilinear equation are
contour integrals along simple closed cycles
$C_\infty$ (for integrals on the left hand side)
and $C_0$ (for integrals on the right hand side)
that encircle the points $z = \infty$ and $z = 0$.
Actually, since these integrals simply extract
the coef\/f\/icient of $z^{-1}$ from Laurent expansion
at those points, we can redef\/ine these integrals
as a genuine linear map from Laurent series to constants:
\begin{gather*}
  \oint\frac{dz}{2\pi i}
    \sum_{n=-\infty}^\infty a_nz^n
  = a_{-1}.
\end{gather*}
As we show below, this bilinear equation
is a generating functional expression of
an inf\/inite number of Hirota equations.

In some cases, it is more convenient
to shift $s$ and $r$ as $s \to s+1$ and  $r \to r+1$.
The outcome is the equation
\begin{gather}
 \oint\frac{dz}{2\pi i}
     z^{s'+r'-s-r-2}e^{\xi(\bst'-\bst,z)}
     \tau(s',r',\bst'-[z^{-1}],\bstbar')
     \tau(s+1,r+1,\bst+[z^{-1}],\bstbar)\nonumber \\
  \qquad\quad{} +
  \oint\frac{dz}{2\pi i}
     z^{s+r-s'-r'-2}e^{\xi(\bst-\bst',z)}
     \tau(s'+1,r'+1,\bst'+[z^{-1}],\bstbar')
     \tau(s,r,\bst-[z^{-1}],\bstbar) \nonumber\\
\qquad{}= \oint\frac{dz}{2\pi i}
     z^{s'-r'-s+r}e^{\xi(\bstbar'-\bstbar,z^{-1})}
     \tau(s'+1,r',\bst',\bstbar'-[z])
     \tau(s,r+1,\bst,\bstbar+[z])\nonumber\\
 \qquad \quad{} +
  \oint\frac{dz}{2\pi i}
     z^{s-r-s'+r'}e^{\xi(\bstbar-\bstbar',z^{-1})}
     \tau(s',r'+1,\bst',\bstbar'+[z])
     \tau(s+1,r,\bst,\bstbar-[z]).
\label{bilin-tau2}
\end{gather}
By changing variables as $z \to z^{-1}$
on the right hand side, this equation can be
converted to a~more symmetric form as
\begin{gather*}
 \oint\frac{dz}{2\pi i}
     z^{s'+r'-s-r-2}e^{\xi(\bst'-\bst,z)}
     \tau(s',r',\bst'-[z^{-1}],\bstbar')
     \tau(s+1,r+1,\bst+[z^{-1}],\bstbar) \nonumber\\
 \qquad\quad{} +
  \oint\frac{dz}{2\pi i}
     z^{s+r-s'-r'-2}e^{\xi(\bst-\bst',z)}
     \tau(s'+1,r'+1,\bst'+[z^{-1}],\bstbar')
     \tau(s,r,\bst-[z^{-1}],\bstbar) \nonumber\\
\qquad{}= \oint\frac{dz}{2\pi i}
     z^{-s'+r'+s-r-2}e^{\xi(\bstbar'-\bstbar,z)}
     \tau(s'+1,r',\bst',\bstbar'-[z^{-1}])
     \tau(s,r+1,\bst,\bstbar+[z^{-1}])\nonumber\\
  \qquad\quad{} +
  \oint\frac{dz}{2\pi i}
     z^{-s+r+s'-r'-2}e^{\xi(\bstbar-\bstbar',z)}
     \tau(s',r'+1,\bst',\bstbar'+[z^{-1}])
     \tau(s+1,r,\bst,\bstbar-[z^{-1}]), \!\!\!
\end{gather*}
though we shall not pursue this line further.

\subsection{Hirota equations}\label{section2.2}

Following the standard procedure,
we now introduce arbitrary constants
\begin{gather*}
  \bsa = (a_1,a_2,\ldots), \qquad
  \bsabar = (\abar_1,\abar_2,\ldots)
\end{gather*}
and shift the continuous variables
$\bst$, $\bst'$, $\bstbar$, $\bstbar'$
in the bilinear equation (\ref{bilin-tau}) as
\begin{gather*}
  \bst' \to \bst - \bsa, \qquad
  \bstbar' \to \bstbar - \bsabar,\qquad
  \bst \to \bst + \bsa, \qquad
  \bstbar \to \bstbar + \bsabar.
\end{gather*}
The bilinear equation thereby takes such a form as
\begin{gather*}
 \oint\frac{dz}{2\pi i}
     z^{s'+r'-s-r}e^{-2\xi(\bsa,z)}
     \tau(s',r',\bst-\bsa-[z^{-1}],\bstbar-\bsabar)
     \tau(s,r,\bst+\bsa+[z^{-1}],\bstbar+\bsabar) \\
 {} +
  \oint\frac{dz}{2\pi i}
     z^{s+r-s'-r'-4}e^{2\xi(\bsa,z)}
     \tau(s'+1,r'+1,\bst-\bsa+[z^{-1}],\bstbar-\bsabar)\\
    \qquad {}\times
     \tau(s-1,r-1,\bst+\bsa-[z^{-1}],\bstbar+\bsabar) \\
= \oint\frac{dz}{2\pi i}
     z^{s'-r'-s+r}e^{-2\xi(\bsabar,z^{-1})}
     \tau(s'+1,r',\bst-\bsa,\bstbar-\bsabar-[z])
     \tau(s-1,r,\bst+\bsa,\bstbar+\bsabar+[z])\\
 {} +
  \oint\frac{dz}{2\pi i}
     z^{s-r-s'+r'}e^{2\xi(\bsabar,z^{-1})}
     \tau(s',r'+1,\bst-\bsa,\bstbar-\bsabar+[z])
     \tau(s,r-1,\bst+\bsa,\bstbar+\bsabar-[z]).
\end{gather*}
With the aid of Hirota's notations
\begin{gather*}
  D_{t_n}f\cdot g = \rd_{t_n}f\cdot g - f\cdot\rd_{t_n}g,\qquad
  D_{\tbar_n}f\cdot g = \rd_{\tbar_n}f\cdot g - f\cdot\rd_{\tbar_n}g,
\end{gather*}
the product of two shifted tau functions
in each term of this equation can be expressed as
\begin{gather*}
   \tau(s',r',\bst-\bsa-[z^{-1}],\bstbar-\bsabar)
   \tau(s,r,\bst+\bsa+[z^{-1}],\bstbar+\bsabar) \\
\qquad  = e^{\xi(\tilde{D}_t,z^{-1})}
    e^{\langle\bsa,D_t\rangle+\langle\bsabar,D_{\tbar}\rangle}
    \tau(s,r,\bst,\bstbar)\cdot\tau(s',r',\bst,\bstbar),
\end{gather*}
etc., where $D_t$ and $D_{\tbar}$ denote the arrays
\begin{gather*}
  D_t = (D_{t_1},D_{t_2},\ldots,D_{t_n},\ldots), \qquad
  D_{\tbar} = (D_{\tbar_1},D_{\tbar_2},\ldots,D_{\tbar_n},\ldots)
\end{gather*}
of Hirota bilinear operators, $\tilde{D}_t$ and $\tilde{D}_{\tbar}$
their variants
\begin{gather*}
  \tilde{D}_t
  = \left(D_{t_1},\frac{1}{2}D_{t_2},\ldots,\frac{1}{n}D_{t_n},\ldots\right), \qquad
  \tilde{D}_{\tbar}
  = \left(D_{\tbar_1},\frac{1}{2}D_{\tbar_2},\ldots,\frac{1}{n}D_{\tbar_n},\ldots\right),
\end{gather*}
and $\langle\bsa,D_t\rangle$ and $\langle\bsabar,D_{\tbar}\rangle$
their linear combinations
\begin{gather*}
  \langle\bsa,D_t\rangle = \sum_{n=1}^\infty a_nD_{t_n}, \qquad
  \langle\bsabar,D_{\tbar}\rangle = \sum_{n=1}^\infty \abar_nD_{\tbar_n}.
\end{gather*}
Let us introduce the functions $h_n(\bst)$, $n \ge 0$,
def\/ined by the generating function
\begin{gather*}
  \sum_{n=0}^\infty h_n(\bst)z^n = e^{\xi(\bst,z)}.
\end{gather*}
The f\/irst few terms read
\begin{gather*}
  h_0(\bst) = 1, \quad h_1(\bst) = t_1,\qquad
  h_2(\bst) = \frac{t_1^2}{2} + t_2,\qquad
  h_3(\bst) = \frac{t_1^3}{6} + t_1t_2 + t_3,\qquad
  \ldots.
\end{gather*}
The prefactors $e^{\pm 2\xi(\bsa,z)}$, etc.,
can be thereby expanded as
\begin{gather*}
  e^{\pm 2\xi(\bsa,z)}
  = \sum_{n=0}^\infty h_n(\pm 2\bsa)z^n, \qquad
  e^{\pm 2\xi(\bsabar,z^{-1})}
  = \sum_{n=0}^\infty h_n(\pm 2\bsabar)z^{-n}.
\end{gather*}
Similarly, the exponential operators
$e^{\pm\xi(\tilde{D}_t,z^{-1})}$, etc.,
can be expanded as
\begin{gather*}
  e^{\pm\xi(\tilde{D}_t,z^{-1})}
  = \sum_{n=0}^\infty h_n(\tilde{D}_t)z^{-n}, \qquad
  e^{\pm\xi(\tilde{D}_{\tbar},z)}
  = \sum_{n=0}^\infty h_n(\tilde{D}_{\tbar})z^n.
\end{gather*}
The bilinear equation thus turns into the Hirota form
\begin{gather}
\sum_{n=0}^\infty h_n(-2\bsa)h_{n+s'+r'-s-r+1}(\tilde{D}_t)
  e^{\langle\bsa,D_t\rangle+\langle\bsabar,D_{\tbar}\rangle}
   \tau(s,r)\cdot\tau(s',r')\nonumber \\
\qquad\quad{}
+ \sum_{n=0}^\infty h_n(2\bsa)h_{n+s+r-s'-r'-3}(-\tilde{D}_t)
    e^{\langle\bsa,D_t\rangle+\langle\bsabar,D_{\tbar}\rangle}
     \tau(s-1,r-1)\cdot\tau(s'+1,r'+1)\nonumber\\
\qquad{}= \sum_{n=0}^\infty h_n(-2\bsabar)h_{n-s'+r'+s-r-1}(\tilde{D}_{\tbar})
    e^{\langle\bsa,D_t\rangle+\langle\bsabar,D_{\tbar}\rangle}
     \tau(s-1,r)\cdot\tau(s'+1,r') \nonumber\\
  \qquad\quad{}
  + \sum_{n=0}^\infty h_n(2\bsabar)h_{n-s+r+s'-r'-1}(-\tilde{D}_{\tbar})
      e^{\langle\bsa,D_t\rangle+\langle\bsabar,D_{\tbar}\rangle}
       \tau(s,r-1)\cdot\tau(s',r'+1).
\label{Hirota-gen}
\end{gather}

The last equation is still a generating functional expression,
from which one can derive an inf\/inite number of equations
by Taylor expansion of both hand sides
at $\bsa = \bszero$ and $\bsabar = \bszero$.
For example, the linear part of the expansion
give the equations
\begin{gather*}
(-2h_{n+s'+r'-s-r+1}(\tilde{D}_t) + h_{s'+r'-s-r+1}(\tilde{D}_t)D_{t_n})
\tau(s,r)\cdot\tau(s',r')\nonumber\\
\qquad\quad{}
+ (2h_{n+s+r-s'-r'-3}(-\tilde{D}_t) + h_{s+r-s'-r'-3}(-\tilde{D}_t)D_{t_n})
  \tau(s-1,r-1)\cdot\tau(s'+1,r'+1)\! \nonumber\\
\qquad{}= h_{-s'+r'+s-r-1}(\tilde{D}_{\tbar})D_{t_n}\tau(s-1,r)\cdot\tau(s'+1,r')\nonumber\\
  \qquad\quad{}
  + h_{-s+r+s'-r'-1}(-\tilde{D}_{\tbar})D_{t_n}\tau(s,r-1)\cdot\tau(s',r'+1)
\end{gather*}
and
\begin{gather}
h_{s'+r'-s-r+1}(\tilde{D}_t)D_{\tbar_n}\tau(s,r)\cdot\tau(s',r') \nonumber\\
\qquad\quad{}+ h_{s+r-s'-r'-3}(-\tilde{D}_t)D_{\tbar_n}\tau(s-1,r-1)\cdot\tau(s'+1,r'+1)\nonumber\\
\qquad{}= (-2h_{n-s'+r'+s-r-1}(\tilde{D}_{\tbar})
   + h_{-s'+r'+s-r-1}(\tilde{D}_{\tbar_n})D_{\tbar_n})
  \tau(s-1,r)\cdot\tau(s'+1,r')\nonumber\\
  \qquad\quad{}
  + (2h_{n-s+r+s'-r'-1}(-\tilde{D}_{\tbar})
     + h_{-s+r+s'-r'-1}(-\tilde{D}_{\tbar})D_{\tbar_n})
    \tau(s,r-1)\cdot\tau(s',r'+1)
\label{Hirota-sp2}
\end{gather}
for $n = 0,1,\ldots$.  In particular,
the special case of (\ref{Hirota-sp2})
where $s' = s$, $r' = r$ and $n = 1$
gives the equation
\begin{gather*}
  \frac{1}{2}D_{t_1}D_{\tbar_1}\tau(s,r)\cdot\tau(s,r)
  + \tau(s-1,r)\tau(s+1,r) - \tau(s,r-1)\tau(s,r+1)
  = 0.
\end{gather*}
Moreover, specializing (\ref{Hirota-gen})
to $\bsa = \bsabar = \bszero$, $s' = s+1$
and $r' = r-1$  yields the equation
\begin{gather}
  D_{t_1}\tau(s,r)\cdot\tau(s+1,r-1)
  + D_{\tbar_1}\tau(s,r-1)\cdot\tau(s+1,r)
  = 0.
\label{Hirota-lowest2}
\end{gather}
These equations give the lowest part of
the whole Hirota equations.

\subsection{Fermionic formula of tau functions}\label{section2.3}

Solutions of these bilinear equations are given
by ground state expectation values of operators
on the Fock space of 2D complex free fermions.

Let us recall basic constituents of the fermion system.
$\psi_j$, $\psi^*_j$ ($j \in \ZZ$) denote
the Fourier modes of fermion f\/ields
\begin{gather*}
  \psi(z) = \sum_{j=-\infty}^\infty \psi_jz^j, \qquad
  \psi^*(z) =\sum_{j=-\infty}^\infty \psi^*_jz^{-j-1}.
\end{gather*}
They obey the anti-commutation relations
\begin{gather*}
  [\psi_j,\psi^*_k]_{+} = \delta_{jk}, \qquad
  [\psi_j,\psi_k]_{+} = [\psi^*_j,\psi^*_k]_{+} = 0.
\end{gather*}
$|0\rangle$ and $\langle 0|$ denote
the vacuum states characterized by
the annihilation conditions
\begin{alignat*}{3}
 & \langle 0|\psi_j = 0 \quad \mbox{for $j \ge 0$},\qquad&&
  \psi_j|0\rangle = 0 \quad \mbox{for $j < 0$},&  \\
 & \langle 0|\psi^*_j = 0 \quad \mbox{for $j < 0$},\qquad &&
  \psi^*_j|0\rangle = 0 \quad \mbox{for $j \ge 0$}.&
\end{alignat*}
The Fock space and its dual space are generated
these vacuum states, and decomposed to
eigenspaces of the charge operator
\begin{gather*}
  H_0 = \sum_{j=-\infty}^\infty {:}\psi_j\psi^*_j{:}
  \qquad (\mbox{normal ordering}).
\end{gather*}
The ground states of the charge-$s$ subspace
are given by
\begin{gather*}
  |s\rangle
  = \begin{cases}
    \psi_{s-1}\cdots\psi_0|0\rangle & \mbox{for $s>0$},\\
    \psi^*_{s}\cdots\psi^*_{-1}|0\rangle & \mbox{for $s<0$},
    \end{cases}  \qquad
  \langle s|
  = \begin{cases}
    \langle 0|\psi^*_0\cdots\psi^*_{s-1} & \mbox{for $s>0$},\\
    \langle 0|\psi_{-1}\cdots\psi_{s} & \mbox{for $s<0$}.
    \end{cases}
\end{gather*}
$H_n$ ($n \in \ZZ$) denote the Fourier modes
\begin{gather*}
  H_n = \sum_{j=-\infty}^\infty{:}\psi_j\psi^*_{j+n}{:}
  \qquad (\mbox{normal ordering})
\end{gather*}
of the $U(1)$ current
\begin{gather*}
  J(z) = {:}\psi(z)\psi^*(z){:}
  = \sum_{n=-\infty}^\infty H_nz^{-n-1}.
\end{gather*}
They obey the commutation relations
\begin{gather*}
  [H_m,H_n] = m\delta_{m+n,0}
\end{gather*}
of a Heisenberg algebra.

Solutions of the bilinear equations are now given by
\begin{gather}
  \tau(s,r,\bst,\bstbar)
  = \langle s+r|e^{H(\bst)}ge^{-\Hbar(\bstbar)}|s-r\rangle,
\label{tau-fermion}
\end{gather}
where $H(\bst)$ and $\Hbar(\bstbar)$ are
the linear combinations
\begin{gather*}
  H(\bst) = \sum_{n=1}^\infty t_nH_n, \qquad
  \Hbar(\bstbar) = \sum_{n=1}^\infty \tbar_nH_{-n}
\end{gather*}
of $H_n$'s, and $g$ is an operator of the form
\begin{gather*}
  g = \exp\left(\sum_{j,k}a_{jk}{:}\psi_j\psi^*_k{:}
       + \sum_{j,k}b_{jk}{:}\psi_j\psi_k{:}
       + \sum_{j,k}c_{jk}{:}\psi^*_j\psi^*_k{:}\right).
\end{gather*}
Note that this operator, unlike $H(\bst)$ and
$\Hbar(\bstbar)$, does not preserve charges,
hence the foregoing expectation value
can take nonzero values for $r \not= 0$.

Let us mention that Willox's original def\/inition
\cite{Willox02,Willox05} of the tau function
is slightly dif\/ferent from (\ref{tau-fermion}).
His def\/inition reads
\begin{gather*}
  \tilde{\tau}(s,r,\bst,\bstbar)
  = \langle s+r|e^{H(t)}e^{\Hbar(\bstbar)}
    ge^{-\Hbar(\bstbar)}|s-r\rangle.
\end{gather*}
This is certainly dif\/ferent from our def\/inition;
for example, Hirota equations are thereby mo\-di\-f\/ied.
The dif\/ference is, however, minimal, because
the two tau functions are connected by
the simple relation
\begin{gather*}
  \tilde{\tau}(s,r,\bst,\bstbar)
  = \exp\left(\sum_{n=1}^\infty nt_n\tbar_n\right)
    \tau(s,r,\bst,\bstbar),
\end{gather*}
so that one can transfer from one def\/inition
to the other freely.

The bilinear equation (\ref{bilin-tau})
is a consequence of the identity
\begin{gather*}
\oint\frac{dz}{2\pi i}
  (\psi(z)g\otimes\psi^*(z)g + \psi^*(z)g\otimes\psi(z)g) \\
\qquad{}= \oint\frac{dz}{2\pi i}
    (g\psi(z)\otimes g\psi^*(z) + g\psi^*(z)\otimes g\psi(z))
\end{gather*}
satisf\/ied by the operator $g$.
This identity implies the equation
\begin{gather*}
\oint\frac{dz}{2\pi i}
  \langle s'+r'+1|e^{H'}\psi(z)ge^{-\Hbar'}|s'-r'\rangle
  \langle s+r-1|e^H\psi^*(z)ge^{-\Hbar}|s-r\rangle \\
\qquad\quad{}
+ \oint\frac{dz}{2\pi i}
    \langle s'+r'+1|e^{H'}\psi^*(z)ge^{-\Hbar'}|s'-r'\rangle
    \langle s+r-1|e^H\psi(z)ge^{-\Hbar}|s-r\rangle\\
\qquad{}= \oint\frac{dz}{2\pi i}
    \langle s'+r'+1|e^{H'}g\psi(z)e^{-\Hbar'}|s'-r'\rangle
    \langle s+r-1|e^Hg\psi^*(z)e^{-\Hbar}|s-r\rangle\\
  \qquad\quad{}
  + \oint\frac{dz}{2\pi i}
      \langle s'+r'+1|e^{H'}g\psi^*(z)e^{-\Hbar'}|s'-r'\rangle
      \langle s+r-1|e^Hg\psi(z)e^{-\Hbar}|s'-r'\rangle,
\end{gather*}
where the abbreviated notations
\begin{gather*}
  H = H(\bst), \qquad
  H' = H(\bst'), \qquad
  \Hbar = \Hbar(\bstbar), \qquad
  \Hbar' = \Hbar(\bstbar')
\end{gather*}
are used.  This equation implies the bilinear equation
(\ref{bilin-tau}) by the bosonization formulae
\begin{gather*}
\langle s|e^{H(\bst)}\psi(z)
   = z^{s-1}e^{\xi(\bst,z)}\langle s-1|e^{H(\bst-[z^{-1}])},\nonumber\\
\langle s|e^{H(\bst)}\psi^*(z)
   = z^{-s-1}e^{-\xi(\bst,z)}\langle s+1|e^{H(\bst+[z^{-1}])}
\end{gather*}
and their duals
\begin{gather*}
\psi(z)e^{-\Hbar(\bstbar)}|s\rangle
   = e^{-\Hbar(\bstbar-[z])}|s+1\rangle z^se^{\xi(\bstbar,z^{-1})},\nonumber\\
\psi^*(z)e^{-\Hbar(\bstbar)}|s\rangle
   = e^{-\Hbar(\bstbar+[z])}|s-1\rangle z^{-s}e^{-\xi(\bstbar,z^{-1})}.
\end{gather*}

\subsection{Relation to DKP hierarchy}\label{section2.4}

The Pfaf\/f--Toda hierarchy contains
an inf\/inite number of copies of
the DKP hierarchy as subsystems.

Such a subsystem shows up by restricting
the variables in (\ref{bilin-tau}) as
\begin{gather*}
  \bstbar' = \bstbar, \qquad
  s = l + r, \qquad s' = l + r',
\end{gather*}
where $l$ is a constant.  (\ref{bilin-tau})
then reduces to the equation
\begin{gather*}
\oint\frac{dz}{2\pi i}
  z^{2r'-2r}e^{\xi(\bst'-\bst,z)}
  \tau(l+r',r',\bst'-[z^{-1}],\bstbar)
  \tau(l+r,r,\bst+[z^{-1}],\bstbar) \\
\qquad{}
 + \oint\frac{dz}{2\pi i}
    z^{2r-2r'-4}e^{\xi(\bst-\bst',z)}
    \tau(l+r'+1,r'+1,\bst'+[z^{-1}],\bstbar)\\
 \qquad\quad{}\times
    \tau(l+r-1,r-1,\bst-[z^{-1}],\bstbar)
= 0,
\end{gather*}
which is substantially the bilinear equation
characterizing tau functions of the DKP hierarchy.
Thus
\begin{gather}
  \tau(l+r,r,\bst,\bstbar)
  = \langle l+2r|e^{H(\bst)}ge^{-\Hbar(\bstbar)}|l\rangle
\label{DKP-tau1}
\end{gather}
turns out to be a tau function of the DKP hierarchy
with respect to $\bst$.

One can derive another family of subsystems
by restricting the variables as
\begin{gather*}
  s = l + 1 - r, \qquad s' = l - 1 - r, \qquad
  \bst' = \bst,
\end{gather*}
where $l$ is a constant.  (\ref{bilin-tau})
thereby reduces to the equation
\begin{gather*}
0 = \oint\frac{dz}{2\pi i}
     z^{-2r'+2r-2}e^{\xi(\bstbar'-\bstbar,z^{-1})}
     \tau(l-r',r',\bst,\bstbar'-[z])
     \tau(l-r,r,\bst,\bstbar+[z]) \\
\phantom{0=}{}
  + \oint\frac{dz}{2\pi i}
    z^{-2r+2r'+2}e^{\xi(\bstbar-\bstbar',z^{-1})}
    \tau(l-1-r',r'+1,\bst,\bstbar'+[z])
    \tau(l+1-r,r-1,\bst,\bstbar-[z]).
\end{gather*}
This is again equivalent to the bilinear equation
for the DKP hierarchy.  Thus
\begin{gather}
  \tau(l-r,r,\bst,\bstbar)
  = \langle l|e^{H(\bst)}ge^{-\Hbar(\bstbar)}|l-2r\rangle
\label{DKP-tau2}
\end{gather}
is a tau function of the DKP hierarchy
with respect to $\bstbar$.

\section{Auxiliary linear problem}\label{section3}

\subsection{Wave functions and dressing operators}\label{section3.1}

To formulate an auxiliary linear problem,
we now introduce the wave functions
\begin{gather}
\Psi_1(s,r,\bst,\bstbar,z)
= z^{s+r}e^{\xi(\bst,z)}
   \frac{\tau(s,r,\bst-[z^{-1}],\bstbar)}{\tau(s,r,\bst,\bstbar)},\nonumber\\
\Psi_2(s,r,\bst,\bstbar,z)
= z^{s+r-2}e^{\xi(\bst,z)}
   \frac{\tau(s-1,r-1,\bst-[z^{-1}],\bstbar)}{\tau(s,r,\bst,\bstbar)},\nonumber\\
\Psi^*_1(s,r,\bst,\bstbar,z)
= z^{-s-r-2}e^{-\xi(\bst,z)}
   \frac{\tau(s+1,r+1,\bst+[z^{-1}],\bstbar)}{\tau(s,r,\bst,\bstbar)},\nonumber\\
\Psi^*_2(s,r,\bst,\bstbar,z)
= z^{-s-r}e^{-\xi(\bst,z)}
   \frac{\tau(s,r,\bst+[z^{-1}],\bstbar)}{\tau(s,r,\bst,\bstbar)}
\label{wave-fn1}
\end{gather}
and their duals
\begin{gather}
\Psibar_1(s,r,\bst,\bstbar,z)
= z^{s-r}e^{\xi(\bstbar,z^{-1})}
   \frac{\tau(s+1,r,\bst,\bstbar-[z])}{\tau(s,r,\bst,\bstbar)},\nonumber\\
\Psibar_2(s,r,\bst,\bstbar,z)
= z^{s-r}e^{\xi(\bstbar,z^{-1})}
   \frac{\tau(s,r-1,\bst,\bstbar-[z])}{\tau(s,r,\bst,\bstbar)},\nonumber\\
\Psibar^*_1(s,r,\bst,\bstbar,z)
= z^{-s+r}e^{-\xi(\bstbar,z^{-1})}
   \frac{\tau(s,r+1,\bst,\bstbar+[z])}{\tau(s,r,\bst,\bstbar)},\nonumber\\
\Psibar^*_2(s,r,\bst,\bstbar,z)
= z^{-s+r}e^{-\xi(\bstbar,z^{-1})}
   \frac{\tau(s-1,r,\bst,\bstbar+[z])}{\tau(s,r,\bst,\bstbar)}.
\label{wave-fn2}
\end{gather}
These wave functions are divided to two groups
with respect to the aforementioned two copies
of the DKP hierarchy.  When the discrete variables $(s,r)$
are restricted on the line $s = l+r$,
the f\/irst four (\ref{wave-fn1}) may be thought of
as wave functions of the DKP hierarchy with
tau function~(\ref{DKP-tau1}).  Similarly,
when $(s,r)$ sit on the line $s = l - r$,
the second four (\ref{wave-fn2}) are to be
identif\/ied with wave functions of the DKP hierarchy
with tau function (\ref{DKP-tau2}).
If the tau function is given by the fermionic formula
(\ref{tau-fermion}), these wave functions, too,
can be written in a fermionic form as
\begin{gather*}
\Psi_1(s,r,\bst,\bstbar,z)
= \frac{\langle s+r+1|e^H\psi(z)ge^{-\Hbar}|s-r\rangle}
   {\langle s+r|e^Hge^{-\Hbar}|s-r\rangle},\nonumber\\
\Psi_2(s,r,\bst,\bstbar,z)
= \frac{\langle s+r-1|e^H\psi(z)ge^{-\Hbar}|s-r\rangle}
   {\langle s+r|e^Hge^{-\Hbar}|s-r\rangle},\nonumber\\
\Psi^*_1(s,r,\bst,\bstbar,z)
= \frac{\langle s+r+1|e^H\psi^*(z)ge^{-\Hbar}|s-r\rangle}
   {\langle s+r|e^Hge^{-\Hbar}|s-r\rangle},\nonumber\\
\Psi^*_2(s,r,\bst,\bstbar,z)
= \frac{\langle s+r-1|e^H\psi^*(z)ge^{-\Hbar}|s-r\rangle}
   {\langle s+r|e^Hge^{-\Hbar}|s-r\rangle},\nonumber
\end{gather*}
and
\begin{gather*}
\Psibar_1(s,r,\bst,\bstbar,z)
= \frac{\langle s+r+1|e^Hg\psi(z)e^{-\Hbar}|s-r\rangle}
   {\langle s+r|e^Hge^{-\Hbar}|s-r\rangle},\nonumber\\
\Psibar_2(s,r,\bst,\bstbar,z)
= \frac{\langle s+r-1|e^Hg\psi(z)e^{-\Hbar}|s-r\rangle}
   {\langle s+r|e^Hge^{-\Hbar}|s-r\rangle},\nonumber\\
\Psibar^*_1(s,r,\bst,\bstbar,z)
= \frac{\langle s+r+1|e^Hg\psi^*(z)e^{-\Hbar}|s-r\rangle}
   {\langle s+r|e^Hge^{-\Hbar}|s-r\rangle},\nonumber\\
\Psibar^*_2(s,r,\bst,\bstbar,z)
= \frac{\langle s+r-1|e^Hg\psi^*(z)e^{-\Hbar}|s-r\rangle}
   {\langle s+r|e^Hge^{-\Hbar}|s-r\rangle}.\nonumber
\end{gather*}

As a consequence of the bilinear equation
(\ref{bilin-tau}), these wave functions
satisfy a system of bilinear equations.
Those equations can be cast into a matrix form as
\begin{gather}
\oint\frac{dz}{2\pi i}
  \Psi_{2\times 2}(s',r',\bst',\bstbar',z)
  \tp{\Psi^*_{2\times 2}(s,r,\bst,\bstbar,z)}\nonumber\\
\qquad{} = \oint\frac{dz}{2\pi i}
    \Psibar_{2\times 2}(s',r',\bst',\bstbar',z)
    \tp{\Psibar^*_{2\times 2}(s,r,\bst,\bstbar,z)},
\label{bilin-Psi}
\end{gather}
where
\begin{gather*}
\Psi_{2\times 2}(s,r,\bst,\bstbar,z)
= \left(\begin{array}{cc}
   \Psi_1(s,r,\bst,\bstbar,z) & \Psi^*_1(s,r,\bst,\bstbar,z)\\
   \Psi_2(s,r,\bst,\bstbar,z) & \Psi^*_2(s,r,\bst,\bstbar,z)
   \end{array}\right),\\
\Psi^*_{2\times 2}(s,r,\bst,\bstbar,z)
= \left(\begin{array}{cc}
   \Psi^*_1(s,r,\bst,\bstbar,z) & \Psi_1(s,r,\bst,\bstbar,z)\\
   \Psi^*_2(s,r,\bst,\bstbar,z) & \Psi_2(s,r,\bst,\bstbar,z)
   \end{array}\right),\\
\Psibar_{2\times 2}(s,r,\bst,\bstbar,z)
= \left(\begin{array}{cc}
   \Psibar_1(s,r,\bst,\bstbar,z) & \Psibar^*_1(s,r,\bst,\bstbar,z)\\
   \Psibar_2(s,r,\bst,\bstbar,z) & \Psibar^*_2(s,r,\bst,\bstbar,z)
   \end{array}\right),\\
\Psibar^*_{2\times 2}(s,r,\bst,\bstbar,z)
= \left(\begin{array}{cc}
   \Psibar^*_1(s,r,\bst,\bstbar,z) & \Psibar_1(s,r,\bst,\bstbar,z)\\
   \Psibar^*_2(s,r,\bst,\bstbar,z) & \Psibar_2(s,r,\bst,\bstbar,z)
   \end{array}\right).
\end{gather*}

Let us now introduce the dressing operators
\begin{alignat*}{3}
 & W_1 = 1 + \sum_{n=1}^\infty w_{1n}e^{-n\rd_s},\qquad&&
  V_1 = \sum_{n=0}^\infty v_{1n}e^{(n+2)\rd_s},& \\
 & W_2 = \sum_{n=0}^\infty w_{2n}e^{-(n+2)\rd_s},\quad&&
  V_2 = 1 + \sum_{n=1}^\infty v_{2n}e^{n\rd_s},& \\
&  \Wbar_1 = \sum_{n=0}^\infty \wbar_{1n}e^{n\rd_s},\qquad &&
  \Vbar_1 = \sum_{n=0}^\infty \vbar_{1n}e^{-n\rd_s}, & \\
&  \Wbar_2 = \sum_{n=0}^\infty \wbar_{2n}e^{n\rd_s},\qquad &&
  \Vbar_2 = \sum_{n=0}^\infty \vbar_{2n}e^{-n\rd_s},&
\end{alignat*}
where $w_{1n}$, etc., are the coef\/f\/icients of
Laurent expansion of the tau-quotient
in the wave functions~(\ref{wave-fn1}) and
(\ref{wave-fn2}), namely,
\begin{alignat*}{3}
& \frac{\tau(s,r,\bst-[z^{-1}],\bstbar)}{\tau(s,r,\bst,\bstbar)}
  = 1 + \sum_{n=1}^\infty w_{1n}z^{-n}, \qquad &&
\frac{\tau(s+1,r+1,\bst+[z^{-1}],\bstbar)}{\tau(s,r,\bst,\bstbar)}
  = \sum_{n=0}^\infty v_{1n}z^{-n}, & \\
& \frac{\tau(s-1,r-1,\bst-[z^{-1}],\bstbar)}{\tau(s,r,\bst,\bstbar)}
  = \sum_{n=0}^\infty w_{2n}z^{-n}, \qquad &&
\frac{\tau(s,r,\bst+[z^{-1}],\bstbar)}{\tau(s,r,\bst,\bstbar)}
  = 1 + \sum_{n=1}^\infty v_{2n}z^{-n}, &
\end{alignat*}
and
\begin{alignat*}{3}
& \frac{\tau(s+1,r,\bst,\bstbar-[z])}{\tau(s,r,\bst,\bstbar)}
  = \sum_{n=0}^\infty \wbar_{1n}z^n, \qquad &&
\frac{\tau(s,r+1,\bst,\bstbar+[z])}{\tau(s,r,\bst,\bstbar)}
  = \sum_{n=0}^\infty \vbar_{1n}z^n, & \\
& \frac{\tau(s,r-1,\bst,\bstbar-[z])}{\tau(s,r,\bst,\bstbar)}
  = \sum_{n=0}^\infty \wbar_{2n}z^n, \qquad &&
\frac{\tau(s-1,r,\bst,\bstbar+[z])}{\tau(s,r,\bst,\bstbar)}
  = \sum_{n=0}^\infty \vbar_{2n}z^n.&
\end{alignat*}
The wave function can be thereby expressed as
\begin{alignat*}{3}
 & \Psi_\alpha(s,r,\bst,\bstbar,z)
    = W_\alpha z^{s+r}e^{\xi(\bst,z)},\qquad &&
  \Psibar_\alpha(s,r,\bst,\bstbar,z)
    = \Wbar_\alpha z^{s-r}e^{\xi(\bstbar,z^{-1})},&\\
 & \Psi^*_\alpha(s,r,\bst,\bstbar,z)
    = V_\alpha z^{-s-r}e^{-\xi(\bst,z)}, \qquad &&
  \Psibar^*_\alpha(s,r,\bst,\bstbar,z)
    = \Vbar_\alpha z^{-s+r}e^{-\xi(\bstbar,z^{-1})}.&
\end{alignat*}

\subsection{Algebraic relations among dressing operators}\label{section3.2}

A technical clue of the following consideration
is a formula that connects wave functions and
dressing operators.  This formula is an analogue
of the formula for the case where the dressing operators
are pseudo-dif\/ferential operators
\cite{DJKM-review,Dickey-book,KvdL93}.
Let us introduce a few notations.
For a~pair of dif\/ference operators of the form
\begin{gather*}
  P = \sum_{n=-\infty}^\infty p_n(s)e^{n\rd_s},\qquad
  Q = \sum_{n=-\infty}^\infty q_n(s)e^{n\rd_s},
\end{gather*}
let $\Psi(s,z)$ and $\Phi(s,z)$ denote the wave functions
\begin{gather*}
  \Psi(s,z) = Pz^s = \sum_{n=-\infty}^\infty p_n(s)z^{n+s},\qquad
  \Phi(s,z) = Qz^{-s} = \sum_{n=-\infty}^\infty q_n(s)z^{-n-s}.
\end{gather*}
Moreover, let $P^*$ denote the formal adjoint
\begin{gather*}
  P^* = \sum_{n=-\infty}^\infty e^{-n\rd_s}p_n(s),
\end{gather*}
and $(P)_{s's}$ the ``matrix elements''
\begin{gather*}
  (P)_{s's} = p_{s-s'}(s').
\end{gather*}
With these notations, the formula reads
\begin{gather}
  \oint\frac{dz}{2\pi i}\Psi(s',z)\Phi(s,z)
  = (Pe^{\rd_s}Q^*)_{s's}
  = (Qe^{-\rd_s}P^*)_{ss'}.
\label{oint-formula}
\end{gather}
One can derive this formula
by straightforward calculations,
which are rather simpler than the case
of pseudo-dif\/ferential operators
\cite{DJKM-review,Dickey-book,KvdL93}.

To illustrate the usage of this formula,
we now derive a set of algebraic relations
satisf\/ied by the dressing operators from
the bilinear equation (\ref{bilin-Psi})
specialized to $\bst' = \bst$ and $\bstbar' = \bstbar$.
Since these relations contain dressing operators
for two dif\/ferent values of $r$,
let us indicate the $(s,r)$ dependence explicitly
as $W(s,r)$, etc.

\begin{theorem}
Specialization of the bilinear equation \eqref{bilin-Psi}
to $\bst' = \bst$ and $\bstbar' = \bstbar$
is equivalent to the algebraic relations
\begin{gather}
W_\alpha(s,r')e^{(r'-r+1)\rd_s}V_\beta(s,r)^*
+ V_\alpha(s,r')e^{(r'-r-1)\rd_s}W_\beta(s,r)^*\nonumber\\
\qquad{}= \Wbar_\alpha(s,r')e^{(r-r'+1)\rd_s}\Vbar_\beta(s,r)^*
  + \Vbar_\alpha(s,r')e^{(r-r'-1)\rd_s}\Wbar_\beta(s,r)^*
\label{WV-relation}
\end{gather}
for $\alpha,\beta = 1,2$.
\end{theorem}

\begin{proof}
The $(1,1)$ component of the specialized
bilinear equation reads
\begin{gather*}
\oint\frac{dz}{2\pi i}\Psi_1(s',r',z)\Psi^*_1(s,r,z)
+ \oint\frac{dz}{2\pi i}\Psi^*_1(s',r',z)\Psi_1(s,r,z)\\
\qquad{}= \oint\frac{dz}{2\pi i}\Psibar_1(s',r',z)\Psibar^*_1(s,r,z)
  + \oint\frac{dz}{2\pi i}\Psibar^*_1(s',r',z)\Psibar_1(s,r,z).
\end{gather*}
By the key formula (\ref{oint-formula}),
each term of this equation can be expressed as
\begin{gather*}
\oint\frac{dz}{2\pi i}\Psi_1(s',r',z)\Psi^*_1(s,r,z)
 = (W_1(s,r')e^{(r'-r+1)\rd_s}V_1(s,r))_{s's},\\
\oint\frac{dz}{2\pi i}\Psi^*_1(s',r',z)\Psi_1(s,r,z)
 = (W_1(s,r)e^{(r-r'+1)\rd_s}V_1(s,r'))_{ss'}\\
\hphantom{\oint\frac{dz}{2\pi i}\Psi^*_1(s',r',z)\Psi_1(s,r,z)}{}
= (V_1(s,r')e^{(r'-r-1)\rd_s}W_1(s,r))_{s's},\\
\oint\frac{dz}{2\pi i}\Psibar_1(s',r',z)\Psibar^*_1(s,r,z)
 = (\Wbar_1(s,r')e^{(r-r'+1)\rd_s}\Vbar_1(s,r))_{s's},\\
\oint\frac{dz}{2\pi i}\Psibar^*_1(s',r',z)\Psibar_1(s,r,z)
 = (\Wbar_1(s,r)e^{(r'-r+1)\rd_s}\Vbar_1(s,r'))_{ss'}\\
\hphantom{\oint\frac{dz}{2\pi i}\Psibar^*_1(s',r',z)\Psibar_1(s,r,z)}{}
= (\Vbar_1(s,r')e^{(r-r'-1)\rd_s}\Wbar_1(s,r))_{s's}.
\end{gather*}
Thus we f\/ind that the $(1,1)$ component
of the specialized bilinear equation
is equivalent to (\ref{WV-relation}) for
$\alpha = \beta = 1$.  The other components
can be treated in the same way.
\end{proof}

In particular, letting $r' = r$ in
(\ref{WV-relation}), we obtain a set of
algebraic relations satisf\/ied by~$W$, $V$, $\Wbar$, $\Vbar$.
We can rewrite these relations
in the following matrix form,
which turns out to be useful later on.
Note that the formal adjoint of
a matrix of operators is def\/ined to be
the transposed matrix of the formal adjoints
of matrix elements as
\begin{gather*}
  \left(\begin{array}{cc}
  A & B\\
  C & D
  \end{array}\right)^*
= \left(\begin{array}{cc}
  A^* & C^*\\
  B^* & D^*
  \end{array}\right).
\end{gather*}

\begin{corollary}
The dressing operators satisfy the algebraic relation
\begin{gather}
  \left(\begin{array}{cc}
  W_1 & \Vbar_1 \\
  W_2 & \Vbar_2
  \end{array}\right)^*
= \left(\begin{array}{cc}
  0 & e^{\rd_s} \\
  -e^{-\rd_s} & 0
  \end{array}\right)
  \left(\begin{array}{cc}
  \Wbar_1 & V_1 \\
  \Wbar_2 & V_2
  \end{array}\right)^{-1}
  \left(\begin{array}{cc}
  0 & -e^{\rd_s} \\
  e^{-\rd_s} & 0
  \end{array}\right)
\label{WV-constraint1}
\end{gather}
or, equivalently,
\begin{gather}
  \left(\begin{array}{cc}
  \Wbar_1 & V_1 \\
  \Wbar_2 & V_2
  \end{array}\right)^*
= \left(\begin{array}{cc}
  0 & e^{\rd_s} \\
  -e^{-\rd_s} & 0
  \end{array}\right)
  \left(\begin{array}{cc}
  W_1 & \Vbar_1 \\
  W_2 & \Vbar_2
  \end{array}\right)^{-1}
  \left(\begin{array}{cc}
  0 & -e^{\rd_s} \\
  e^{-\rd_s} & 0
  \end{array}\right).
\label{WV-constraint2}
\end{gather}
\end{corollary}

\begin{proof}
Let us examine (\ref{WV-relation}) in the case
where $r' = r$.  The $(1,1)$ component reads
\begin{gather*}
W_1 e^{\rd_s}V_1^* + V_1 e^{-\rd_s}W_1^*
= \Wbar_1 e^{\rd_s}\Vbar_1^* + \Vbar_1 e^{-\rd_s}\Wbar_1^*.
\end{gather*}
Among the four terms in this relation,
$W_1e^{\rd_s}V^*_1$ and $\Vbar_1e^{-\rd_s}\Wbar^*_1$
are linear combinations of $e^{-\rd_s},e^{-2\rd_s},\ldots$,
and $V_1e^{-\rd_s}W^*_1$ and $\Wbar_1e^{\rd_s}\Vbar^*_1$
are linear combinations of $e^{\rd_s},e^{2\rd_s},\ldots$.
Therefore this relation splits into
the two relations
\begin{gather*}
  V_1e^{-\rd_s}W^*_1 = \Wbar_1e^{\rd_s}\Vbar^*_1,\qquad
  W_1e^{\rd_s}V^*_1 = \Vbar_1e^{-\rd_s}\Wbar^*_1,
\end{gather*}
which are actually equivalent. In the same way,
we can derive the relations
\begin{gather*}
  V_2e^{-\rd_s}W^*_2 = \Wbar_2e^{\rd_s}\Vbar^*_2,\qquad
  W_2e^{\rd_s}V^*_2 = \Vbar_2e^{-\rd_s}\Wbar^*_2
\end{gather*}
from the $(2,2)$ component of (\ref{WV-relation}).
Let us now consider the $(1,2)$ component,
which we rewrite as
\begin{gather*}
  W_1e^{\rd_s}V^*_2 - \Vbar_1e^{-\rd_s}\Wbar^*_2
  = \Wbar_1e^{\rd_s}\Vbar^*_2 - V_1e^{-\rd_s}W^*_2.
\end{gather*}
The left hand side is a sum of $e^{\rd_s}$
and a linear combination of $1,e^{-\rd_s},\ldots$,
and the right hand side is a sum of $e^{\rd_s}$
and a linear combination of $e^{2\rd_s},e^{3\rd_s},\ldots$.
Therefore both hand sides should be equal to $e^{\rd_s}$,
namely,
\begin{gather*}
  W_1e^{\rd_s}V^*_2 - \Vbar_1e^{-\rd_s}\Wbar^*_2
  = \Wbar_1e^{\rd_s}\Vbar^*_2 - V_1e^{-\rd_s}W^*_2
  = e^{\rd_s}.
\end{gather*}
By the same reasoning, we can derive the relations
\begin{gather*}
  W_2e^{\rd_s}V^*_1 - \Vbar_2e^{-\rd_s}\Wbar^*_1
  = \Wbar_2e^{\rd_s}\Vbar^*_1 - V_2e^{-\rd_s}W^*_1
  = - e^{-\rd_s}
\end{gather*}
from the $(2,1)$ component of (\ref{WV-relation}).
These relations can be cast into a matrix form as
(\ref{WV-constraint1}) and~(\ref{WV-constraint2}).
\end{proof}

(\ref{WV-constraint1}) and (\ref{WV-constraint2})
may be thought of as constraints preserved
under time evolutions with respect to $\bst$ and
$\bstbar$.  Actually, the discrete variable~$r$, too,
has to be interpreted as a time variable.
Letting $r' = r + 1$ in (\ref{WV-relation}),
we can see how the dressing operators evolve in~$r$.

\begin{corollary}
The dressing operators with $r$ shifted by one
are related to the unshifted dressing operators as
\begin{gather}
  \left(\begin{array}{cc}
  W_1(s,r+1) & \Vbar_1(s,r+1)\\
  W_2(s,r+1) & \Vbar_2(s,r+1)
  \end{array}\right)
  \left(\begin{array}{cc}
  e^{\rd_s} & 0\\
  0 & e^{-\rd_s}
  \end{array}\right)
= \left(\begin{array}{cc}
  A & B\\
  C & 0
  \end{array}\right)
  \left(\begin{array}{cc}
  W_1 & \Vbar_1\\
  W_2 & \Vbar_2
  \end{array}\right),
\nonumber\\
  \left(\begin{array}{cc}
  \Wbar_1(s,r+1) & V_1(s,r+1)\\
  \Wbar_2(s,r+1) & V_2(s,r+1)
  \end{array}\right)
  \left(\begin{array}{cc}
  e^{-\rd_s} & 0\\
  0 & e^{\rd_s}
  \end{array}\right)
= \left(\begin{array}{cc}
  A & B\\
  C & 0
  \end{array}\right)
  \left(\begin{array}{cc}
  \Wbar_1 & V_1\\
  \Wbar_2 & V_2
  \end{array}\right),
\label{WV-evolve-r}
\end{gather}
where
\begin{gather}
  A = e^{\rd_s}
    + \left(\log\frac{\tau(s+1,r)}{\tau(s,r+1)}\right)_{t_1}
    + \frac{\tau(s+1,r+1)\tau(s-1,r)}{\tau(s,r+1)\tau(s,r)}e^{-\rd_s},\nonumber\\
  B = - \frac{\tau(s+1,r+1)}{\tau(s,r)}e^{\rd_s},\qquad
  C = \frac{\tau(s-1,r)}{\tau(s,r+1)}e^{-\rd_s}.
\label{ABC}
\end{gather}
\end{corollary}

\begin{proof}
When $r' = r + 1$, (\ref{WV-relation}) reads
\begin{gather*}
  W_\alpha(s,r+1)e^{2\rd_s}V^*_\beta
  + V_\alpha(s,r+1)W^*_\beta
  = \Wbar_\alpha(s,r+1)\Vbar^*_\beta
    + \Vbar_\alpha(s,r+1)e^{-2\rd_s}\Wbar^*_\beta.
\end{gather*}
These equations can be cast into
a matrix form as
\begin{gather*}
  \left(\begin{array}{cc}
  W_1(s,r+1) & \Vbar_1(s,r+1)\\
  W_2(s,r+1) & \Vbar_2(s,r+1)
  \end{array}\right)
  \left(\begin{array}{cc}
  0 & e^{2\rd_s}\\
  - e^{-2\rd_s} & 0
  \end{array}\right)
  \left(\begin{array}{cc}
  \Wbar^*_1 & \Wbar^*_2\\
  V^*_1 & V^*_2
  \end{array}\right) \\
\qquad{}= \left(\begin{array}{cc}
  \Wbar_1(s,r+1) & V_1(s,r+1)\\
  \Wbar_2(s,r+1) & V_2(s,r+1)
  \end{array}\right)
  \left(\begin{array}{cc}
  0 & 1\\
  -1 & 0
  \end{array}\right)
  \left(\begin{array}{cc}
  W^*_1 & W^*_2\\
  \Vbar^*_1 & \Vbar^*_2
  \end{array}\right).
\end{gather*}
Noting that
\begin{gather*}
  \left(\begin{array}{cc}
  \Wbar^*_1 & \Wbar^*_2\\
  V^*_1 & V^*_2
  \end{array}\right)
 =\left(\begin{array}{cc}
  \Wbar_1 & V_1\\
  \Wbar_2 & V_2
  \end{array}\right)^*,
\qquad
  \left(\begin{array}{cc}
  W^*_1 & W^*_2\\
  \Vbar^*_1 & \Vbar^*_2
  \end{array}\right)
 =\left(\begin{array}{cc}
  W_1 & \Vbar_1\\
  W_2 & \Vbar_2
  \end{array}\right)^*,
\end{gather*}
we can use (\ref{WV-constraint1})
and (\ref{WV-constraint2}) to rewrite
this equation as
\begin{gather*}
  \left(\begin{array}{cc}
  W_1(s,r+1) & \Vbar_1(s,r+1)\\
  W_2(s,r+1) & \Vbar_2(s,r+1)
  \end{array}\right)
  \left(\begin{array}{cc}
  e^{\rd_s} & 0\\
  0 & e^{-\rd_s}
  \end{array}\right)
  \left(\begin{array}{cc}
  W_1 & \Vbar_1\\
  W_2 & \Vbar_2
  \end{array}\right)^{-1}\\
\qquad{}= \left(\begin{array}{cc}
  \Wbar_1(s,r+1) & V_1(s,r+1)\\
  \Wbar_2(s,r+1) & V_2(s,r+1)
  \end{array}\right)
  \left(\begin{array}{cc}
  e^{-\rd_s} & 0\\
  0 & e^{\rd_s}
  \end{array}\right)
  \left(\begin{array}{cc}
  \Wbar_1 & V_1\\
  \Wbar_2 & V_2
  \end{array}\right)^{-1}.
\end{gather*}
By the def\/inition of the dressing operators,
the left hand side of this equation
is a matrix of operators of the form
\begin{gather*}
  \left(\begin{array}{cc}
  e^{\rd_s} + w_{11}(s,r+1) - w_{11}(s+1,r) + \cdots
  & - \dfrac{\vbar_{10}(s+1,r)}{\vbar_{20}(s+1,r)}e^{\rd_s}
    + \cdots \\
  \bullet e^{-\rd_s} + \bullet e^{-2\rd_s} + \cdots
  & \bullet e^{-\rd_s} + \bullet e^{-2\rd_s} + \cdots
  \end{array}\right)
\end{gather*}
($\bullet$ denotes a function),
and the right hand side take such a form as
\begin{gather*}
  \left(\begin{array}{cc}
  \dfrac{\wbar_{10}(s,r+1)}{\wbar_{10}(s-1,r)}e^{-\rd_s} + \cdots
  & \bullet e^{\rd_s} + \bullet e^{2\rd_s} + \cdots\\
  \dfrac{\wbar_{20}(s,r+1)}{\wbar_{10}(s-1,r)}e^{-\rd_s} + \cdots
  & \bullet e^{\rd_s} + \bullet e^{2\rd_s} + \cdots
  \end{array}\right).
\end{gather*}
Consequently,
\begin{gather*}
  \left(\begin{array}{cc}
  W_1(s,r+1) & \Vbar_1(s,r+1)\\
  W_2(s,r+1) & V_2(s,r+1)
  \end{array}\right)
  \left(\begin{array}{cc}
  e^{\rd_s} & 0\\
  0 & e^{-\rd_s}
  \end{array}\right)
  \left(\begin{array}{cc}
  W_1 & \Vbar_1\\
  W_2 & \Vbar_2
  \end{array}\right)^{-1}
= \left(\begin{array}{cc}
  A & B\\
  C & 0
  \end{array}\right)
\end{gather*}
and
\begin{gather*}
\left(\begin{array}{cc}
  \Wbar_1(s,r+1) & V_1(s,r+1)\\
  \Wbar_2(s,r+1) & V_2(s,r+1)
  \end{array}\right)
  \left(\begin{array}{cc}
  e^{-\rd_s} & 0\\
  0 & e^{\rd_s}
  \end{array}\right)
  \left(\begin{array}{cc}
  \Wbar_1 & V_1\\
  \Wbar_2 & V_2
  \end{array}\right)^{-1}
= \left(\begin{array}{cc}
  A & B\\
  C & 0
  \end{array}\right),
\end{gather*}
where
\begin{gather*}
 A = e^{\rd_s} + w_{11}(s,r+1) - w_{11}(s+1,r)
    + \frac{\wbar_{10}(s,r+1)}{\wbar_{10}(s-1,r)}e^{-\rd_s},\\
 B = - \frac{\vbar_{10}(s+1,r)}{\vbar_{20}(s+1,r)}e^{\rd_s},\qquad
 C = w_{20}(s,r+1)e^{-\rd_s}
    = \frac{\wbar_{20}(s,r+1)}{\wbar_{10}(s-1,r)}e^{-\rd_s}.
\end{gather*}
Rewriting this result in terms of the tau functions,
we obtain the formulae (\ref{ABC}) of $A$, $B$, $C$.
\end{proof}

\subsection{Evolution equations of dressing operators}\label{section3.3}

The dressing operators turn out to satisfy
a set of evolution equations with respect to
the continuous time variables $\bst$ and $\bstbar$ as well.
To present the result, let us introduce the notations
\begin{gather*}
  (P)_{\ge 0} = \sum_{n\ge 0} p_ne^{n\rd_s},\qquad
 (P)_{>0} = \sum_{n>0} p_ne^{n\rd_s},\\
  (P)_{\le 0} = \sum_{n\le 0} p_ne^{n\rd_s},\qquad
 (P)_{<0} = \sum_{n<0} p_ne^{n\rd_s}
\end{gather*}
of truncated operators for dif\/ference operators
of the form
\begin{gather*}
  P = \sum_{n=-\infty}^\infty p_ne^{n\rd_s}.
\end{gather*}

\begin{theorem}
The dressing operators satisfy the equations
\begin{gather}
  \left(\begin{array}{cc}
  W_{1,t_n} + W_1e^{n\rd_s} & \Vbar_{1,t_n}\\
  W_{2,t_n} + W_2e^{n\rd_s} & \Vbar_{2,t_n}
  \end{array}\right)
 =\left(\begin{array}{cc}
  A_n & B_n\\
  C_n & D_n
  \end{array}\right)
  \left(\begin{array}{cc}
  W_1 & \Vbar_1\\
  W_2 & \Vbar_2
  \end{array}\right),
\nonumber\\
  \left(\begin{array}{cc}
  \Wbar_{1,t_n} & V_{1,t_n} - V_1e^{-n\rd_s}\\
  \Wbar_{2,t_n} & V_{2,t_n} - V_2e^{-n\rd_s}
  \end{array}\right)
 =\left(\begin{array}{cc}
  A_n & B_n\\
  C_n & D_n
  \end{array}\right)
  \left(\begin{array}{cc}
  \Wbar_1 & V_1\\
  \Wbar_2 & V_2
  \end{array}\right),
\label{WV-evolve-t}
\end{gather}
where the subscript $t_n$ in $W_{1,t_n}$, etc.,
stands for differentiating the coefficients
of the difference operators by $t_n$,
\begin{gather*}
  W_{\alpha,t_n} = \frac{\rd W_\alpha}{\rd t_n}, \qquad
  V_{\alpha,t_n} = \frac{\rd V_\alpha}{\rd t_n}, \qquad
  \Wbar_{\alpha,t_n} = \frac{\rd\Wbar_\alpha}{\rd t_n},\qquad
  \Vbar_{\alpha,t_n} = \frac{\rd\Vbar_\alpha}{\rd t_n},
\end{gather*}
and $A_n$, $B_n$, $C_n$ and $D_n$ are defined as
\begin{gather*}
  A_n = (W_1e^{(n+1)\rd_s}V_2^*e^{-\rd_s})_{\ge 0}
         + (V_1e^{-(n+1)\rd_s}W_2^*e^{-\rd_s})_{<0},\\
  B_n = - (W_1e^{(n+1)\rd_s}V_1^*e^{\rd_s})_{>0}
         - (V_1e^{-(n+1)\rd_s}W_1^*e^{\rd_s})_{\le 0},\\
  C_n = (W_2e^{(n+1)\rd_s}V_2^*e^{-\rd_s})_{\ge 0}
         + (V_2e^{-(n+1)\rd_s}W_2^*e^{-\rd_s})_{<0},\\
  D_n = - (W_2e^{-(n+1)\rd_s}V_1^*e^{\rd_s})_{>0}
         - (V_2e^{-(n+1)\rd_s}W_1^*e^{\rd_s})_{\le 0}.
\end{gather*}
\end{theorem}

\begin{proof}
Dif\/ferentiate the bilinear equation (\ref{bilin-Psi})
by $t'_n$ and specialize the variables to
$\bst' = \bst$, $\bstbar' = \bstbar$,
$s' = s$ and $r' = r$.  This leads to the equation
\begin{gather*}
\oint\frac{dz}{2\pi i}\rd_{t_n}\Psi_{2\times 2}(s,r,z)
  \cdot\tp{\Psi^*_{2\times 2}(s,r,z)}
= \oint\frac{dz}{2\pi i}\rd_{t_n}\Psibar^*_{2\times 2}(s,r,z)
    \cdot\tp{\Psibar^*_{2\times 2}(s,r,z)}.
\end{gather*}
The $t_n$-derivatives of the wave functions
in this equation can be expressed as
\begin{gather*}
\rd_{t_n}\Psi_\alpha(s,r,z)
  = (W_{\alpha,t_n} + W_\alpha e^{n\rd_s})z^{s+r}e^{\xi(\bst,z)},\\
\rd_{t_n}\Psi^*_\alpha(s,r,z)
  = (V_{\alpha,t_n} - V_\alpha e^{-n\rd_s})z^{-s-r}e^{-\xi(\bst,z)},\\
\rd_{t_n}\Psibar_\alpha(s,r,z)
  = \Wbar_{\alpha,t_n}z^{s-r}e^{\xi(\bstbar,z^{-1})},\\
\rd_{t_n}\Psibar^*_\alpha(s,r,z)
  = \Vbar_{\alpha,t_n}e^{-s+r}e^{-\xi(\bstbar,z^{-1})}.
\end{gather*}
We now use the key formula (\ref{oint-formula})
to convert the last equation to equations
for the dressing operators.
Those equations can be cast into a matrix form as
\begin{gather*}
\left(\begin{array}{cc}
W_{1,t_n} + W_1e^{n\rd_s} & \Vbar_{1,t_n}\\
W_{2,t_n} + W_2e^{n\rd_s} & \Vbar_{1,t_n}
\end{array}\right)
\left(\begin{array}{cc}
0 & - e^{\rd_s}\\
e^{-\rd_s} & 0
\end{array}\right)
\left(\begin{array}{cc}
\Wbar_1^* & \Wbar_2^*\\
V_1^* & V_2^*
\end{array}\right) \\
\qquad{}= \left(\begin{array}{cc}
  \Wbar_{1,t_n} & V_{1,t_n} - V_1e^{-n\rd_s}\\
  \Wbar_{2,t_n} & V_{2,t_n} - V_2e^{-n\rd_s}
  \end{array}\right)
  \left(\begin{array}{cc}
  0 & - e^{\rd_s}\\
  e^{-\rd_s} & 0
  \end{array}\right)
  \left(\begin{array}{cc}
  W_1^* & W_2^*\\
  \Vbar_1^* & \Vbar_2^*
  \end{array}\right).
\end{gather*}
By (\ref{WV-constraint1}) and (\ref{WV-constraint2}),
this equation can be rewritten as
\begin{gather*}
\left(\begin{array}{cc}
W_{1,t_n} + W_1e^{n\rd_s} & \Vbar_{1,t_n}\\
W_{2,t_n} + W_2e^{n\rd_s} & \Vbar_{1,t_n}
\end{array}\right)
\left(\begin{array}{cc}
W_1 & \Vbar_1\\
W_2 & \Vbar_2
\end{array}\right)^{-1}\\
\qquad{}= \left(\begin{array}{cc}
  \Wbar_{1,t_n} & V_{1,t_n} - V_1e^{-n\rd_s}\\
  \Wbar_{2,t_n} & V_{2,t_n} - V_2e^{-n\rd_s}
  \end{array}\right)
  \left(\begin{array}{cc}
  \Wbar_1 & V_1\\
  \Wbar_2 & V_2
  \end{array}\right)^{-1}.
\end{gather*}
Let $A_n$, $B_n$, $C_n$, $D_n$ denote the matrix elements
of both hand sides of this equation.
To complete the proof, we have to show
that these dif\/ference operators do have
the form stated in the theorem.
To this end, let us compare the two
dif\/ferent expressions
\begin{gather*}
  \left(\begin{array}{cc}
  A_n & B_n\\
  C_n & D_n
  \end{array}\right)
 =\left(\begin{array}{cc}
  W_{1,t_n} + W_1e^{n\rd_s} & \Vbar_{1,t_n}\\
  W_{2,t_n} + W_2e^{n\rd_s} & \Vbar_{2,t_n}
  \end{array}\right)
  \left(\begin{array}{cc}
  W_1 & \Vbar_1\\
  W_2 & \Vbar_2
  \end{array}\right)^{-1}\\
\hphantom{\left(\begin{array}{cc}
  A_n & B_n\\
  C_n & D_n
  \end{array}\right)}{}
  = \left(\begin{array}{cc}
  W_{1,t_n} + W_1e^{n\rd_s} & \Vbar_{1,t_n}\\
  W_{2,t_n} + W_2e^{n\rd_s} & \Vbar_{2,t_n}
  \end{array}\right)
  \left(\begin{array}{cc}
  e^{\rd_s}V_2^*e^{-\rd_s} & - e^{\rd_s}V_1^*e^{\rd_s}\\
  - e^{-\rd_s}\Wbar_2^*e^{-\rd_s} & e^{-\rd_s}\Wbar_1^*e^{\rd_s}
  \end{array}\right)
\end{gather*}
and
\begin{gather*}
  \left(\begin{array}{cc}
  A_n & B_n\\
  C_n & D_n
  \end{array}\right)
=\left(\begin{array}{cc}
  \Wbar_{1,t_n} & V_{1,t_n} - V_1e^{-n\rd_s}\\
  \Wbar_{2,t_n} & V_{2,t_n} - V_2e^{-n\rd_s}
  \end{array}\right)
  \left(\begin{array}{cc}
  \Wbar_1 & V_1\\
  \Wbar_2 & V_2
  \end{array}\right)^{-1}\\
\hphantom{\left(\begin{array}{cc}
  A_n & B_n\\
  C_n & D_n
  \end{array}\right)}{}
= \left(\begin{array}{cc}
  \Wbar_{1,t_n} & V_{1,t_n} - V_1e^{-n\rd_s}\\
  \Wbar_{2,t_n} & V_{2,t_n} - V_2e^{-n\rd_s}
  \end{array}\right)
  \left(\begin{array}{cc}
  e^{\rd_s}\Vbar_2^*e^{-\rd_s} & - e^{\rd_s}\Vbar_1^*e^{\rd_s}\\
  - e^{-\rd_s}W_2^*e^{-\rd_s} & e^{-\rd_s}W_1^*e^{\rd_s}
  \end{array}\right)
\end{gather*}
of the matrix of these operators
that can be derived from the foregoing
construction and the algebraic relations
(\ref{WV-constraint1}) and (\ref{WV-constraint2}).
As regards~$A_n$, this implies that
\begin{gather*}
A_n = (W_{1,t_n} + W_1e^{n\rd_s})e^{\rd_s}V_2^*e^{-\rd_s}
       - \Vbar_{1,t_n}e^{-\rd_s}\Wbar_2^*e^{-\rd_s}\\
\phantom{A_n}{} = \Wbar_{1,t_n}e^{\rd_s}\Vbar_2^*e^{-\rd_s}
       - (V_{1,t_n} - V_1e^{-n\rd_s})e^{-\rd_s}W_2^*e^{-\rd_s}.
\end{gather*}
From the $( \ )_{\ge 0}$ part of the f\/irst line,
we have the identity
\begin{gather*}
  (A_n)_{\ge 0} = (W_1e^{(n+1)\rd_s}V_2^*e^{-\rd_s})_{\ge 0},
\end{gather*}
and from the $( \ )_{<0}$ part of the second line,
similarly,
\begin{gather*}
  (A_n)_{<0} = (V_1e^{-(n+1)\rd_s}W_2^*e^{-\rd_s})_{<0}.
\end{gather*}
Thus $A_n$ turns out to be given by the sum of
these operators as
\begin{gather*}
  A_n = (W_1e^{(n+1)\rd_s}V_2^*e^{-\rd_s})_{\ge 0}
        + (V_1e^{-(n+1)\rd_s}W_2^*e^{-\rd_s})_{<0}.
\end{gather*}
The other operators $B_n$, $C_n$, $D_n$, too, can be
identif\/ied in a fully parallel manner.
\end{proof}

In much the same way, the following evolution equations
in $\bstbar$ can be derived.

\begin{theorem}
The dressing operators satisfy the equations
\begin{gather}
  \left(\begin{array}{cc}
  W_{1,\tbar_n} & \Vbar_{1,\tbar_n} - \Vbar_ne^{n\rd_s}\\
  W_{2,\tbar_n} & \Vbar_{2,\tbar_n} - \Vbar_2e^{n\rd_s}
  \end{array}\right)
 =\left(\begin{array}{cc}
  \Abar_n & \Bbar_n\\
  \Cbar_n & \Dbar_n
  \end{array}\right)
  \left(\begin{array}{cc}
  W_1 & \Vbar_1\\
  W_2 & \Vbar_2
  \end{array}\right),
\nonumber\\
  \left(\begin{array}{cc}
  \Wbar_{1,\tbar_n} + \Wbar_1e^{-n\rd_s} & V_{1,\tbar_n}\\
  \Wbar_{2,\tbar_n} + \Wbar_2e^{-n\rd_s} & V_{2,\tbar_n}
  \end{array}\right)
=\left(\begin{array}{cc}
  \Abar_n & \Bbar_n\\
  \Cbar_n & \Dbar_n
  \end{array}\right)
  \left(\begin{array}{cc}
  \Wbar_1 & V_1\\
  \Wbar_2 & V_2
  \end{array}\right),
\label{WV-evolve-tbar}
\end{gather}
where
\begin{gather*}
  \Abar_n  = (\Wbar_1e^{-(n-1)\rd_s}\Vbar_2^*e^{-\rd_s})_{<0}
           + (\Vbar_1e^{(n-1)\rd_s}\Wbar_2^*e^{-\rd_s})_{\ge 0},\\
  \Bbar_n  = - (\Wbar_1e^{-(n-1)\rd_s}\Vbar_1^*e^{\rd_s})_{\le 0}
             - (\Vbar_1e^{(n-1)\rd_s}\Wbar_1^*e^{\rd_s})_{>0},\\
  \Cbar_n  = (\Wbar_2e^{-(n-1)\rd_s}\Vbar_2^*e^{-\rd_s})_{<0}
           + (\Vbar_2e^{(n-1)\rd_s}\Wbar_2^*e^{-\rd_s})_{\ge 0},\\
  \Dbar_n  = - (\Wbar_2e^{-(n-1)\rd_s}\Vbar_1^*e^{\rd_s})_{\le 0}
             - (\Vbar_2e^{(n-1)\rd_s}\Wbar_1^*e^{\rd_s})_{> 0}.
\end{gather*}
\end{theorem}

\subsection{Auxiliary linear equations}\label{section3.4}

The evolution equations (\ref{WV-evolve-t}),
(\ref{WV-evolve-tbar}) and (\ref{WV-evolve-r})
for the dressing operators can be readily cast
into auxiliary linear equations for the wave functions.

\begin{corollary}
The wave functions satisfy the following
linear equations:
\begin{gather}
  e^{\rd_r}\left(\begin{array}{cccc}
  \Psi_1 & \Psi^*_1 & \Psibar_1 & \Psibar^*_1\\
  \Psi_2 & \Psi^*_2 & \Psibar_2 & \Psibar^*_2
  \end{array}\right)
 = \left(\begin{array}{cc}
  A & B \\
  C & 0
  \end{array}\right)
  \left(\begin{array}{cccc}
  \Psi_1 & \Psi^*_1 & \Psibar_1 & \Psibar^*_1\\
  \Psi_2 & \Psi^*_2 & \Psibar_2 & \Psibar^*_2
  \end{array}\right),
\label{lin-Psi-r}
\\
  \rd_{t_n}\left(\begin{array}{cccc}
  \Psi_1 & \Psi^*_1 & \Psibar_1 & \Psibar^*_1\\
  \Psi_2 & \Psi^*_2 & \Psibar_2 & \Psibar^*_2
  \end{array}\right)
 = \left(\begin{array}{cc}
  A_n & B_n \\
  C_n & D_n
  \end{array}\right)
  \left(\begin{array}{cccc}
  \Psi_1 & \Psi^*_1 & \Psibar_1 & \Psibar^*_1\\
  \Psi_2 & \Psi^*_2 & \Psibar_2 & \Psibar^*_2
  \end{array}\right),
\label{lin-Psi-t}
\\
  \rd_{\tbar_n}\left(\begin{array}{cccc}
  \Psi_1 & \Psi^*_1 & \Psibar_1 & \Psibar^*_1\\
  \Psi_2 & \Psi^*_2 & \Psibar_2 & \Psibar^*_2
  \end{array}\right)
 = \left(\begin{array}{cc}
  \Abar_n & \Bbar_n \\
  \Cbar_n & \Dbar_n
  \end{array}\right)
  \left(\begin{array}{cccc}
  \Psi_1 & \Psi^*_1 & \Psibar_1 & \Psibar^*_1\\
  \Psi_2 & \Psi^*_2 & \Psibar_2 & \Psibar^*_2
  \end{array}\right).
\label{lin-Psi-tbar}
\end{gather}
\end{corollary}

Note that each of (\ref{lin-Psi-t}),
(\ref{lin-Psi-tbar}) and (\ref{lin-Psi-r})
is a collective expression of four sets
of linear equations, namely,
\begin{gather*}
  \left(\begin{array}{c}
  e^{\rd_r}\Phi_1\\
  e^{\rd_r}\Phi_2
  \end{array}\right)
=\left(\begin{array}{cc}
  A & B \\
  C & 0
  \end{array}\right)
  \left(\begin{array}{c}
  \Phi_1\\
  \Phi_2
  \end{array}\right),
\qquad
  \left(\begin{array}{c}
  \rd_{t_n}\Phi_1\\
  \rd_{t_n}\Phi_2
  \end{array}\right)
=\left(\begin{array}{cc}
  A_n & B_n \\
  C_n & D_n
  \end{array}\right)
  \left(\begin{array}{c}
  \Phi_1\\
  \Phi_2
  \end{array}\right),
\\
  \left(\begin{array}{c}
  \rd_{\tbar_n}\Phi_1\\
  \rd_{\tbar_n}\Phi_2
  \end{array}\right)
=\left(\begin{array}{cc}
  \Abar_n & \Bbar_n \\
  \Cbar_n & \Dbar_n
  \end{array}\right)
  \left(\begin{array}{c}
  \Phi_1\\
  \Phi_2
  \end{array}\right)
\end{gather*}
for the four pairs $\Phi_\alpha = \Psi_\alpha,
\Psi^*_\alpha, \Psibar_\alpha, \Psibar^*_\alpha$
($\alpha = 1,2$) of wave functions.
The lowest ($n = 1$) equations of (\ref{lin-Psi-t})
and (\ref{lin-Psi-tbar}) agree with Willox's result
\cite{Willox02,Willox05}:
\begin{gather}
  \left(\begin{array}{cc}
  \rd_{t_1}\Phi_1(s,r,\mu)\\
  \rd_{t_1}\Phi_2(s,r,\mu)
  \end{array}\right)\nonumber\\
\qquad{}= \left(\begin{array}{cc}
    e^{\rd_s} + \left(\log\frac{\tau(s+1,r)}{\tau(s,r)}\right)_{t_1}&
      - \frac{\tau(s+1,r+1)}{\tau(s,r)}e^{\rd_s}\\
    \frac{\tau(s-1,r-1)}{\tau(s,r)}e^{-\rd_s}&
      - e^{-\rd_s} + \left(\log\frac{\tau(s-1,r)}{\tau(s,r)}\right)_{t_1}
  \end{array}\right)
  \left(\begin{array}{cc}
    \Phi_1(s,r,\mu)\\
    \Phi_2(s,r,\mu)
  \end{array}\right),
\label{lin-Psi-t1}
\\
  \left(\begin{array}{cc}
  \rd_{\tbar_1}\Phi_1(s,r,\mu)\\
  \rd_{\tbar_1}\Phi_2(s,r,\mu)
  \end{array}\right)\nonumber\\
\qquad{}= \left(\begin{array}{cc}
    \frac{\tau(s+1,r)\tau(s-1,r)}{\tau(s,r)^2}e^{-\rd_s}&
      - \frac{\tau(s+1,r)\tau(s,r+1)}{\tau(s,r)^2}e^{\rd_s}\\
    \frac{\tau(s-1,r)\tau(s,r-1)}{\tau(s,r)^2}e^{-\rd_s}&
      - \frac{\tau(s+1,r)\tau(s-1,r)}{\tau(s,r)^2}e^{\rd_s}
  \end{array}\right)
  \left(\begin{array}{cc}
    \Phi_1(s,r,\mu)\\
    \Phi_2(s,r,\mu)
  \end{array}\right).
\label{lin-Psi-tbar1}
\end{gather}

Thus we have obtained auxiliary linear equations
for the Pfaf\/f--Toda hierarchy.  Apart from the fact
that dif\/ference operators play a central role,
this auxiliary linear problem resembles that of
the DKP hierarchy (see Appendix).   This is a manifestation
of the common Lie algebraic structure \cite{JM83,KvdL98}
that underlies these hierarchies.

Let us specify an algebraic structure
in the building blocks of the auxiliary linear problem.
Let $U$, $\Ubar$, $P_n$, $\Pbar_n$ and $J$
denote the matrix operators
\begin{gather*}
U
  = \left(\begin{array}{cc}
    W_1 & \Vbar_1\\
    W_2 & \Vbar_2
    \end{array}\right),\qquad \Ubar
  = \left(\begin{array}{cc}
    \Wbar_1 & V_1\\
    \Wbar_2 & V_2
    \end{array}\right),\qquad P_n
  = \left(\begin{array}{cc}
    A_n & B_n\\
    C_n & D_n
    \end{array}\right),\\
     \Pbar_n
  = \left(\begin{array}{cc}
    \Abar_n & \Bbar_n\\
    \Cbar_n & \Dbar_n
    \end{array}\right),\qquad
 J = \left(\begin{array}{cc}
      0 & e^{\rd_s}\\
      - e^{-\rd_s} & 0
      \end{array}\right),\qquad J^{-1} = - J.
\end{gather*}
With these notations, (\ref{WV-constraint1})
can be rewritten as
\begin{gather*}
  U^* = J\Ubar^{-1}J^{-1},\qquad
  \Ubar^* = JU^{-1}J^{-1}.
\end{gather*}
This exhibits a Lie group structure
(now realized in terms of dif\/ference operators).
Moreover, (\ref{WV-evolve-t}) and (\ref{WV-evolve-tbar})
imply that $P_n$ and $\Pbar_n$ can be expressed as
\begin{gather*}
  P_n
 = U_{t_n}U^{-1}
   + U\left(\begin{array}{cc}
     e^{n\rd_s} & 0 \\
     0 & 0
     \end{array}\right)U^{-1} = \Ubar_{t_n}\Ubar^{-1}
   + \Ubar\left(\begin{array}{cc}
     0 & 0\\
     0 & - e^{-n\rd_s}
     \end{array}\right)\Ubar^{-1},\\
  \Pbar_n
 = U_{\tbar_n}U^{-1}
   + U\left(\begin{array}{cc}
     0 & 0\\
     0 & - e^{n\rd_s}
     \end{array}\right)U^{-1} = \Ubar_{\tbar_n}\Ubar^{-1}
   + \Ubar\left(\begin{array}{cc}
     e^{-n\rd_s} & 0\\
     0 & 0
     \end{array}\right)\Ubar^{-1}.
\end{gather*}
We can conf\/irm by straightforward
calculations that $P_n$ and $\Pbar_n$
satisfy the algebraic relations
\begin{gather*}
  P_n^* = - JP_nJ^{-1},\qquad
  \Pbar_n^* = - J\Pbar_nJ^{-1},
\end{gather*}
which are obviously a Lie algebraic version
of the foregoing constraints for $U$ and $\Ubar$.
In components, these relations read
\begin{gather*}
  A_n^* = - e^{\rd_s}D_ne^{-\rd_s}, \qquad
  B_n^* = e^{-\rd_s}B_ne^{-\rd_s},\qquad
  C_n^* = e^{\rd_s}C_ne^{\rd_s}, \qquad
  D_n^* = - e^{-\rd_s}A_ne^{\rd_s},
\end{gather*}
and
\begin{gather*}
  \Abar_n^* = - e^{\rd_s}\Dbar_ne^{-\rd_s}, \qquad
  \Bbar_n^* = e^{-\rd_s}\Bbar_ne^{-\rd_s},\qquad
  \Cbar_n^* = e^{\rd_s}\Cbar_ne^{\rd_s}, \qquad
  \Dbar_n^* = - e^{-\rd_s}\Abar_ne^{\rd_s}.
\end{gather*}
These relations are parallel to algebraic relations
in the case of the DKP hierarchy (see Appendix).

\section{Dif\/ference Fay identities}\label{section4}

\subsection{How to derive dif\/ference Fay identities}\label{section4.1}

We now derive six Fay-like identities
with parameters $\lambda$ and $\mu$
from the bilinear equation (\ref{bilin-tau2})
by specializing the free variables therein as follows:
\begin{itemize}\itemsep=0pt
\item[1a)]$\bst' = \bst+[\lambda^{-1}]+[\mu^{-1}]$,
$\bstbar' = \bstbar$, $s' = s+1$, $r' = r$;
\item[1b)]$\bst' = \bst+[\lambda^{-1}]+[\mu^{-1}]$,
$\bstbar' = \bstbar$, $s' = s$, $r' = r+1$;
\item[2a)]$\bst' = \bst$, $\bstbar' = \bstbar+[\lambda]+[\mu]$,
$s' = s-1$, $r' = r$;
\item[2b)]$\bst' = \bst$, $\bstbar' = \bstbar+[\lambda]+[\mu]$,
$s' = s$, $r' = r+1$;
\item[3a)]$\bst' = \bst+[\lambda^{-1}]$,
$\bstbar' = \bstbar + [\mu]$, $s' = s$, $r' = r$;
\item[3b)]$\bst' = \bst+[\lambda^{-1}]$,
$\bstbar' = \bstbar + [\mu]$, $s' = s-1$, $r' = r+1$.
\end{itemize}
To clarify the meaning of calculations, we now
assume that the integrals in (\ref{bilin-tau2})
are contour integrals along simple closed curves
$C_\infty$ and $C_0$ encircling the points $z = \infty$
and $z = 0$ respectively.  $\lambda$ and $\mu$
are understood to be in a particular position
specif\/ied below.

\paragraph*{1a) and 1b).}
$\lambda$ and $\mu$ are assumed to sit
on the far side (closer to $z = \infty$)
of the contour $C_\infty$.
The exponential factors in the integrand
thereby become rational functions as
\begin{gather*}
  e^{\xi(\bst'-\bst,z)}
  = \frac{1}{(1-z/\lambda)(1-z/\mu)}, \qquad
  e^{\xi(\bstbar'-\bstbar,z^{-1})} = 1.
\end{gather*}
Thus the bilinear equation (\ref{bilin-tau2})
reduce to the equations
\begin{gather*}
\oint_{C_\infty}\frac{dz}{2\pi i}
     \frac{\lambda\mu}{z(z-\lambda)(z-\mu)}
     \tau(s+1,r,\bst+[\lambda^{-1}]+[\mu^{-1}]-[z^{-1}],\bstbar)
     \tau(s+1,r+1,\bst+[z^{-1}],\bstbar) \\
\quad {}+ \oint_{C_\infty}\frac{dz}{2\pi i}
     \frac{(z-\lambda)(z-\mu)}{z^3\lambda\mu}
     \tau(s+2,r+1,\bst+[\lambda^{-1}]+[\mu^{-1}]+[z^{-1}],\bstbar)
     \tau(s,r,\bst-[z^{-1}],\bstbar) \\
= \oint_{C_0}\frac{dz}{2\pi i}z
     \tau(s+2,r,\bst+[\lambda^{-1}]+[\mu^{-1}],\bstbar-[z])
     \tau(s,r+1,\bst,\bstbar+[z])\\
 \quad{}+ \oint_{C_0}\frac{dz}{2\pi i}\frac{1}{z}
     \tau(s+1,r+1,\bst+[\lambda^{-1}]+[\mu^{-1}],\bstbar+[z])
     \tau(s+1,r,\bst,\bstbar-[z])
\end{gather*}
in the case of 1a) and
\begin{gather*}
\oint_{C_\infty}\frac{dz}{2\pi i}
     \frac{\lambda\mu}{z(z-\lambda)(z-\mu)}
     \tau(s,r+1,\bst+[\lambda^{-1}]+[\mu^{-1}]-[z^{-1}],\bstbar)
     \tau(s+1,r+1,\bst+[z^{-1}],\bstbar) \\
\quad{}+ \oint_{C_\infty}\frac{dz}{2\pi i}
     \frac{(z-\lambda)(z-\mu)}{z^3\lambda\mu}
     \tau(s+1,r+2,\bst+[\lambda^{-1}]+[\mu^{-1}]+[z^{-1}],\bstbar)
     \tau(s,r,\bst-[z^{-1}],\bstbar) \\
=\oint_{C_0}\frac{dz}{2\pi i}\frac{1}{z}
     \tau(s+1,r+1,\bst+[\lambda^{-1}]+[\mu^{-1}],\bstbar-[z])
     \tau(s,r+1,\bst,\bstbar+[z])\\
 \quad{}+ \oint_{C_0}\frac{dz}{2\pi i}z
     \tau(s,r+2,\bst+[\lambda^{-1}]+[\mu^{-1}],\bstbar+[z])
     \tau(s+1,r,\bst,\bstbar-[z])
\end{gather*}
in the case of 1b).  The contour integrals in these equation
can be calculated by residue calculus.  For example,
the f\/irst integral on the left hand side is given
by the sum of residues of the integrand at $z = \lambda,\mu$;
the other contour integrals can be treated in the same way.
The outcome are the equations
\begin{gather}
  - \frac{\mu}{\lambda-\mu}\tau(s+1,r,\bst+[\mu^{-1}],\bstbar)
    \tau(s+1,r+1,\bst+[\lambda^{-1}],\bstbar)\nonumber\\
  \qquad {}- \frac{\lambda}{\mu-\lambda}
    \tau(s+1,r,\bst+[\lambda^{-1}],\bstbar)
    \tau(s+1,r+1,\bst+[\mu^{-1}],\bstbar)\nonumber\\
   \qquad {}+ \frac{1}{\lambda\mu}
    \tau(s+2,r+1,\bst+[\lambda^{-1}]+[\mu^{-1}],\bstbar)
    \tau(s,r,\bst,\bstbar)\nonumber\\
  \qquad = \tau(s+1,r+1,\bst+[\lambda^{-1}]+[\mu^{-1}],\bstbar)
    \tau(s+1,r,\bst,\bstbar),
\label{dFay1a}
\\
  - \frac{\mu}{\lambda-\mu}\tau(s,r+1,\bst+[\mu^{-1}],\bstbar)
    \tau(s+1,r+1,\bst+[\lambda^{-1}],\bstbar)\nonumber\\
  \qquad{}- \frac{\lambda}{\mu-\lambda}
    \tau(s,r+1,\bst+[\lambda^{-1}],\bstbar)
    \tau(s+1,r+1,\bst+[\mu^{-1}],\bstbar)\nonumber\\
  \qquad{}+ \frac{1}{\lambda\mu}
    \tau(s+1,r+2,\bst+[\lambda^{-1}]+[\mu^{-1}],\bstbar)
    \tau(s,r,\bst,\bstbar)\nonumber\\
   \qquad= \tau(s+1,r+1,\bst+[\lambda^{-1}]+[\mu^{-1}],\bstbar)
    \tau(s,r+1,\bst,\bstbar).
\label{dFay1b}
\end{gather}

\paragraph*{2a) and 2b).}
$\lambda$ and $\mu$ are assumed to be inside $C_0$
(nearer to $z = 0$).  The bilinear equation~(\ref{bilin-tau2})
turn into the equations
\begin{gather*}
\oint_{C_\infty}\frac{dz}{2\pi i}\frac{1}{z^3}
     \tau(s-1,r,\bst-[z^{-1}],\bstbar+[\lambda]+[\mu])
     \tau(s+1,r+1,\bst+[z^{-1}],\bstbar) \\
\qquad\quad{}+ \oint_{C_\infty}\frac{dz}{2\pi i}\frac{1}{z}
     \tau(s,r+1,\bst+[z^{-1}],\bstbar+[\lambda]+[\mu])
     \tau(s,r,\bst-[z^{-1}],\bstbar) \\
\qquad=\oint_{C_0}\frac{dz}{2\pi i}\frac{z}{(z-\lambda)(z-\mu)}
     \tau(s,r,\bst,\bstbar+[\lambda]+[\mu]-[z])
     \tau(s,r+1,\bst,\bstbar+[z])\\
 \qquad\quad{}+ \oint_{C_0}\frac{dz}{2\pi i}
     \frac{(z-\lambda)(z-\mu)}{z}
     \tau(s-1,r+1,\bst,\bstbar+[\lambda]+[\mu]+[z])
     \tau(s+1,r,\bst,\bstbar-[z])
\end{gather*}
in the case of 2a) and
\begin{gather*}
\oint_{C_\infty}\frac{dz}{2\pi i}\frac{1}{z}
     \tau(s,r+1,\bst-[z^{-1}],\bstbar+[\lambda]+[\mu])
     \tau(s+1,r+1,\bst+[z^{-1}],\bstbar) \\
\qquad\quad{}+ \oint_{C_\infty}\frac{dz}{2\pi i}\frac{1}{z^3}
     \tau(s+1,r+2,\bst+[z^{-1}],\bstbar+[\lambda]+[\mu])
     \tau(s,r,\bst-[z^{-1}],\bstbar) \\
\qquad{}=\oint_{C_0}\frac{dz}{2\pi i}\frac{z}{(z-\lambda)(z-\mu)}
     \tau(s+1,r+1,\bst,\bstbar+[\lambda]+[\mu]-[z])
     \tau(s,r+1,\bst,\bstbar+[z])\\
 \qquad\quad{}+ \oint_{C_0}\frac{dz}{2\pi i}
     \frac{(z-\lambda)(z-\mu)}{z}
     \tau(s,r+2,\bst,\bstbar+[\lambda]+[\mu]+[z])
     \tau(s+1,r,\bst,\bstbar-[z])
\end{gather*}
in the case of 2b).  By residue calculus,
we obtain the equations
\begin{gather}
  \tau(s,r+1,\bst,\bstbar+[\lambda]+[\mu])\tau(s,r,\bst,\bstbar)\nonumber\\
  \qquad = \frac{\lambda}{\lambda-\mu}\tau(s,r,\bst,\bstbar+[\mu])
    \tau(s,r+1,\bst,\bstbar+[\lambda])\nonumber\\
  \qquad{}+ \frac{\mu}{\mu-\lambda}
    \tau(s,r,\bst,\bstbar+[\lambda])\tau(s,r+1,\bst,\bstbar+[\mu])\nonumber\\
  \qquad{}+ \lambda\mu\tau(s-1,r+1,\bst,\bstbar+[\lambda]+[\mu])
    \tau(s+1,r,\bst,\bstbar),
\label{dFay2a}
\\
  \tau(s,r+1,\bst,\bstbar+[\lambda]+[\mu])\tau(s+1,r+1,\bst,\bstbar)\nonumber\\
 \qquad = \frac{\lambda}{\lambda-\mu}\tau(s+1,r+1,\bst,\bstbar+[\mu])
    \tau(s,r+1,\bst,\bstbar+[\lambda])\nonumber\\
  \qquad{}+ \frac{\mu}{\mu-\lambda}
    \tau(s+1,r+1,\bst,\bstbar+[\lambda])\tau(s,r+1,\bst,\bstbar+[\mu])\nonumber\\
  \qquad{}+ \lambda\mu\tau(s,r+2,\bst,\bstbar+[\lambda]+[\mu])
    \tau(s+1,r,\bst,\bstbar).
\label{dFay2b}
\end{gather}

\paragraph*{3a) and 3b).}
$\lambda$ and $\mu$ are assumed to be
on the far side of $C_\infty$ and inside $C_0$
respectively.  The bilinear equation~(\ref{bilin-tau2})
turn into the equations
\begin{gather*}
\oint_{C_\infty}\frac{dz}{2\pi i}\frac{-\lambda}{z^2(z-\lambda)}
  \tau(s,r,\bst+[\lambda^{-1}]-[z^{-1}],\bstbar+[\mu])
  \tau(s+1,r+1,\bst+[z^{-1}],\bstbar)\\
\qquad\quad{}+ \oint_{C_\infty}\frac{dz}{2\pi i}
  \frac{(z-\lambda)}{-z^2\lambda}
  \tau(s+1,r+1,\bst+[\lambda^{-1}]+[z^{-1}],\bstbar+[\mu])
  \tau(s,r,\bst-[z^{-1}],\bstbar)\\
  \qquad{}= \oint_{C_0}\frac{dz}{2\pi i}\frac{z}{z-\mu}
    \tau(s+1,r,\bst+[\lambda^{-1}],\bstbar+[\mu]-[z])
    \tau(s,r+1,\bst,\bstbar+[z])\\
  \qquad\quad{}+ \oint_{C_0}\frac{dz}{2\pi i}\frac{z-\mu}{z}
    \tau(s,r+1,\bst+[\lambda^{-1}],\bstbar+[\mu]+[z])
    \tau(s+1,r,\bst,\bstbar-[z])
\end{gather*}
in the case of 3a) and
\begin{gather*}
\oint_{C_\infty}\frac{dz}{2\pi i}\frac{-\lambda}{z^2(z-\lambda)}
  \tau(s-1,r+1,\bst+[\lambda^{-1}]-[z^{-1}],\bstbar+[\mu])
  \tau(s+1,r+1,\bst+[z^{-1}],\bstbar)\\
\qquad\quad{}+ \oint_{C_\infty}\frac{dz}{2\pi i}
  \frac{(z-\lambda)}{-z^2\lambda}
  \tau(s,r+2,\bst+[\lambda^{-1}]+[z^{-1}],\bstbar+[\mu])
  \tau(s,r,\bst-[z^{-1}],\bstbar)\\
  \qquad{}= \oint_{C_0}\frac{dz}{2\pi i}\frac{1}{z(z-\mu)}
    \tau(s,r+1,\bst+[\lambda^{-1}],\bstbar+[\mu]-[z])
    \tau(s,r+1,\bst,\bstbar+[z])\\
  \qquad\quad{}+ \oint_{C_0}\frac{dz}{2\pi i}z(z-\mu)
    \tau(s-1,r+2,\bst+[\lambda^{-1}],\bstbar+[\mu]+[z])
    \tau(s+1,r,\bst,\bstbar-[z])
\end{gather*}
in the case of 3b), and boil down to the equations
\begin{gather}
  \lambda^{-1}\tau(s,r,\bst,\bstbar+[\mu])
  \tau(s+1,r+1,\bst+[\lambda^{-1}],\bstbar)\nonumber\\
  \qquad{}- \lambda^{-1}
  \tau(s+1,r+1,\bst+[\lambda^{-1}],\bstbar+[\mu])
  \tau(s,r,\bst,\bstbar) \nonumber\\
  \qquad{}= \mu\tau(s+1,r,\bst+[\lambda^{-1}],\bstbar)
    \tau(s,r+1,\bst,\bstbar+[\mu])\nonumber\\
  \qquad{}- \mu\tau(s,r+1,\bst+[\lambda^{-1}],\bstbar+[\mu])
    \tau(s+1,r,\bst,\bstbar),
\label{dFay3a}
\\
  \lambda^{-1}\tau(s-1,r+1,\bst,\bstbar+[\mu])
  \tau(s+1,r+1,\bst+[\lambda^{-1}],\bstbar)\nonumber\\
  \qquad{}
- \lambda^{-1}\tau(s,r+2,\bst+[\lambda^{-1}],\bstbar+[\mu])
  \tau(s,r,\bst,\bstbar) \nonumber\\
  \qquad= \mu^{-1}\tau(s,r+1,\bst+[\lambda^{-1}],\bstbar)
    \tau(s,r+1,\bst,\bstbar+[\mu])\nonumber\\
  \qquad{}- \mu^{-1}\tau(s,r+1,\bst+[\lambda^{-1}],\bstbar+[\mu])
    \tau(s,r+1,\bst,\bstbar).
\label{dFay3b}
\end{gather}

Actually, these calculations are meaningful
even if the contour integrals are understood
to be genuine algebraic operators that extract
the coef\/f\/icient of $z^{-1}$ from Laurent series.
Thus we are led to the following conclusion:

\begin{theorem}
The bilinear equation \eqref{bilin-tau} implies
the Fay-like identities \eqref{dFay1a}--\eqref{dFay3b}.
\end{theorem}

In the rest of this paper, (\ref{dFay1a})--(\ref{dFay3b})
are referred to as ``dif\/ference Fay identities''.
The structure of these Fay-like equations is
similar to the dif\/ference Fay identities
of the Toda hierarchy \cite{Zabrodin01,Teo06,Takasaki07},
though the latter are three-term relations and
given on a $1D$ lattice.

\subsection{Relation to auxiliary linear problem}\label{section4.2}

We now show that the dif\/ference Fay identities
are closely related to the auxiliary linear equations.
To this end, let us rewrite the identities
in the language of the wave functions as follows.

\begin{theorem}
The difference Fay identities \eqref{dFay1a}--\eqref{dFay3b} are equivalent to the system of
the following four equations:
\begin{gather}
e^{-D(\lambda)}\Phi_1(s,r,\mu) + \lambda^{-1}\Phi_1(s+1,r,\mu)
- \frac{e^{-D(\lambda)}\tau(s+1,r)/\tau(s+1,r)}
    {e^{-D(\lambda)}\tau(s,r)/\tau(s,r)}\Phi_1(s,r,\mu)\nonumber\\
\qquad{} - \lambda^{-1}\frac{e^{-D(\lambda)}\tau(s+1,r)/\tau(s+1,r)}
    {e^{-D(\lambda)}\tau(s,r)/\tau(s+1,r+1)}
    e^{-D(\lambda)}\Phi_2(s+1,r,\mu)
= 0,
\label{lin-gen1}
\\
e^{D(\lambda)}\Phi_2(s,r,\mu) + \lambda^{-1}\Phi_2(s-1,r,\mu)
- \frac{e^{D(\lambda)}\tau(s-1,r)/\tau(s-1,r)}
    {e^{D(\lambda)}\tau(s,r)/\tau(s,r)}\Phi_2(s,r,\mu)\nonumber\\
\qquad{}- \lambda^{-1}\frac{e^{D(\lambda)}\tau(s-1,r)/\tau(s-1,r)}
    {e^{D(\lambda)}\tau(s,r)/\tau(s-1,r-1)}
    e^{D(\lambda)}\Phi_1(s-1,r,\mu)
= 0,
\label{lin-gen2}
\\
e^{-\Dbar(\lambda)}\Phi_1(s,r,\mu) - \Phi_1(s,r,\mu)
+ \lambda\frac{e^{-\Dbar(\lambda)}\tau(s+1,r)/\tau(s,r)}
    {e^{-\Dbar(\lambda)}\tau(s,r)/\tau(s-1,r)}\Phi_1(s-1,r,\mu)\nonumber\\
\qquad{}- \lambda\frac{e^{-\Dbar(\lambda)}\tau(s+1,r)/\tau(s,r)}
    {e^{-\Dbar(\lambda)}\tau(s,r)/\tau(s,r+1)}
    e^{-\Dbar(\lambda)}\Phi_2(s+1,r,\mu)
= 0,
\label{lin-gen3}
\\
e^{\Dbar(\lambda)}\Phi_2(s,r,\mu) - \Phi_2(s,r,\mu)
+ \lambda\frac{e^{\Dbar(\lambda)}\tau(s-1,r)/\tau(s,r)}
    {e^{\Dbar(\lambda)}\tau(s,r)/\tau(s+1,r)}\Phi_2(s+1,r,\mu)\nonumber\\
\qquad{}- \lambda\frac{e^{\Dbar(\lambda)}\tau(s-1,r)/\tau(s,r)}
    {e^{\Dbar(\lambda)}\tau(s,r)/\tau(s,r-1)}
    e^{\Dbar(\lambda)}\Phi_1(s-1,r,\mu)
= 0
\label{lin-gen4}
\end{gather}
for the four pairs $\Phi_\alpha = \Psi_\alpha, \Psi^*_\alpha,
\Psibar_\alpha, \Psibar^*_\alpha$ $(\alpha = 1,2)$
of wave functions, where $D(z)$ and $\Dbar(z)$ denote
the differential operators
\begin{gather*}
  D(z) = \sum_{n=1}^\infty\frac{z^{-n}}{n}\rd_{t_n}, \qquad
  \Dbar(z) = \sum_{n=1}^\infty\frac{z^n}{n}\rd_{\tbar_n}.
\end{gather*}
\end{theorem}

\begin{proof}
One can derive the four dif\/ference Fay identities
from (\ref{lin-gen1})--(\ref{lin-gen4})
by straightforward calculations.  Actually,
this turns out to be largely redundant,
namely, each dif\/ference Fay identity appears
more than once while processing the twelve equations
of (\ref{lin-gen1})--(\ref{lin-gen4}).
Nevertheless the calculations on the whole,
are reversible, proving the converse simultaneously.
Since the whole calculations are considerably lengthy,
let us demonstrate it by deriving (\ref{dFay1a})
and (\ref{dFay1b}) from (\ref{lin-gen1}) and
(\ref{lin-gen2}) for $\Phi_\alpha = \Psi_\alpha$
($\alpha = 1,2$); the other cases are fully parallel.
Recall that $\Psi_\alpha$'s can be expressed as
\begin{gather*}
  \Psi_1(s,r,z)
   = \frac{\tau(s,r,\bst-[z^{-1}],\bstbar)}
     {\tau(s,r,\bst,\bstbar)}z^{s+r}e^{\xi(\bst,z)},\\
  \Psi_2(s,r,z)
   = \frac{\tau(s-1,r-1,\bst-[z^{-1}],\bstbar)}
     {\tau(s,r,\bst,\bstbar)}z^{s+r-2}e^{\xi(\bst,z)}.
\end{gather*}
(\ref{lin-gen1}) thereby reads
\begin{gather*}
e^{-D(\lambda)}\Psi_1(s,r,\mu) + \lambda^{-1}\Psi_1(s+1,r,\mu)
- \frac{e^{-D(\lambda)}\tau(s+1,r)/\tau(s+1,r)}
    {e^{-D(\lambda)}\tau(s,r)/\tau(s,r)}\Psi_1(s,r,\mu)\\
\qquad{} - \lambda^{-1}\frac{e^{-D(\lambda)}\tau(s+1,r)/\tau(s+1,r)}
    {e^{-D(\lambda)}\tau(s,r)/\tau(s+1,r+1)}
    e^{-D(\lambda)}\Psi_2(s+1,r,\mu)
= 0.
\end{gather*}
Noting the identity
\begin{gather*}
  e^{-D(\lambda)}e^{\xi(\bst,\mu)}
  = (1 - \mu/\lambda)e^{\xi(\bst,\mu)},
\end{gather*}
we thus obtain the equation
\begin{gather*}
\left(1 - \frac{\mu}{\lambda}\right)
\frac{\tau(s,r,\bst-[\lambda^{-1}]-[\mu^{-1}],\bstbar)}
     {\tau(s,r,\bst-[\lambda^{-1}],\bstbar)}
+ \frac{\mu}{\lambda}
  \frac{\tau(s+1,r,\bst-[\mu^{-1}],\bstbar)}
       {\tau(s+1,r,\bst,\bstbar)}\\
\qquad{}- \frac{\tau(s+1,r,\bst-[\lambda^{-1}],\bstbar)
               \tau(s,r,\bst-[\mu^{-1}],\bstbar)}
              {\tau(s+1,r,\bst,\bstbar)
               \tau(s,r,\bst-[\lambda^{-1}],\bstbar)}\\
\qquad{}- \frac{1}{\lambda\mu}\left(1 - \frac{\mu}{\lambda}\right)
         \frac{\tau(s+1,r+1,\bst,\bstbar)
               \tau(s,r-1,\bst-[\lambda^{-1}]-[\mu^{-1}],\bstbar)}
              {\tau(s+1,r,\bst,\bstbar)
               \tau(s,r,\bst-[\mu^{-1}],\bstbar)}
= 0
\end{gather*}
for the tau function.  Rewriting this equation as
\begin{gather*}
\tau(s,r,\bst-[\lambda^{-1}]-[\mu^{-1}],\bstbar)\tau(s+1,r,\bst,\bstbar)
+ \frac{\mu}{\lambda-\mu}
         \tau(s+1,r,\bst-[\mu^{-1}],\bstbar)
         \tau(s,r,\bst-[\lambda^{-1}],\bstbar)\\
\qquad{}- \frac{\lambda}{\lambda-\mu}
         \tau(s+1,r,\bst-[\lambda^{-1}],\bstbar)
         \tau(s,r,\bst-[\mu^{-1}],\bstbar)\\
\qquad{}- \frac{1}{\lambda\mu}\tau(s+1,r+1,\bst,\bstbar)
         \tau(s,r-1,\bst-[\lambda^{-1}]-[\mu^{-1}],\bstbar)
= 0
\end{gather*}
and shifting the variables as
$\bst \to \bst + [\lambda^{-1}]+[\mu^{-1}]$
and $r \to r+1$, we arrive at (\ref{dFay1b}).
Let us now consider (\ref{lin-gen2}), namely,
\begin{gather*}
e^{D(\lambda)}\Psi_2(s,r,\mu) + \lambda^{-1}\Psi_2(s-1,r,\mu)
- \frac{e^{D(\lambda)}\tau(s-1,r)/\tau(s-1,r)}
    {e^{D(\lambda)}\tau(s,r)/\tau(s,r)}\Psi_2(s,r,\mu)\\
\qquad{}- \lambda^{-1}\frac{e^{D(\lambda)}\tau(s-1,r)/\tau(s-1,r)}
    {e^{D(\lambda)}\tau(s,r)/\tau(s-1,r-1)}
    e^{D(\lambda)}\Psi_1(s-1,r,\mu)
= 0.
\end{gather*}
This equation turns into the equation
\begin{gather*}
\frac{\lambda}{\lambda-\mu}
\tau(s-1,r-1,\bst+[\lambda^{-1}]-[\mu^{-1}],\bstbar)
\tau(s-1,r,\bst,\bstbar)\\
\qquad{}+ \frac{1}{\lambda\mu}
         \tau(s-2,r-1,\bst-[\mu^{-1}],\bstbar)
         \tau(s,r,\bst+[\lambda^{-1}],\bstbar)\\
\qquad{}- \tau(s-1,r,\bst+[\lambda^{-1}],\bstbar)
         \tau(s-1,r-1,\bst-[\mu^{-1}],\bstbar)\\
\qquad{}- \frac{\mu}{\lambda-\mu}
         \tau(s-1,r-1,\bst,\bstbar)
         \tau(s-1,r,\bst+[\lambda^{-1}]-[\mu^{-1}],\bstbar)
= 0
\end{gather*}
for the tau function, and upon shifting the variables
as $\bst \to \bst + [\mu^{-1}]$, $s \to s + 2$
and $r \to r + 1$, reduces to (\ref{dFay1a}).
It will be obvious that these calculations are reversible.
\end{proof}

Expanded in powers of $\lambda$,
(\ref{lin-gen1})--(\ref{lin-gen4})
generate an inf\/inite set of linear equations
for the wave functions $\Phi_\alpha(s,r,\mu)$.
As we show below, these linear equations
are equivalent to the auxiliary linear equations
(\ref{lin-Psi-t}) and (\ref{lin-Psi-tbar}).
Thus (\ref{lin-gen1})--(\ref{lin-gen4})
turn out to give a generating functional expression
of these auxiliary linear equations.

Since the four equations (\ref{lin-gen1})--(\ref{lin-gen4})
can be treated in the same manner, let us illustrate
the calculations in the case of (\ref{lin-gen1}).
Among the four terms in this equation,
the f\/irst and fourth terms can be readily expanded
by the identity
\begin{gather*}
  e^{-D(\lambda)}
  = \sum_{n=0}^\infty h_n(-\tilde{\rd}_t)\lambda^{-n}, \qquad
  \tilde{\rd}_t
  = \left(\rd_{t_1},\frac{1}{2}\rd_{t_2},\ldots,
     \frac{1}{n}\rd_{t_n},\ldots\right).
\end{gather*}
As regards the coef\/f\/icients of the second and third terms,
we can rewrite them as
\begin{gather*}
\frac{e^{-D(\lambda)}\tau(s+1,r)/\tau(s+1,r)}
     {e^{-D(\lambda)}\tau(s,r)/\tau(s,r)}
= 1
   - \left(\log\frac{\tau(s+1,r)}{\tau(s,r)}\right)_{t_1}\lambda^{-1}
   + f_2\lambda^{-2} + f_3\lambda^{-3} + \cdots
\end{gather*}
and
\begin{gather*}
\frac{e^{-D(\lambda)}\tau(s+1,r)/\tau(s+1,r)}
     {e^{-D(\lambda)}\tau(s,r)/\tau(s+1,r+1)}\\
\qquad= \frac{\tau(s+1,r+1)}{\tau(s,r)}
  \left(1
    - \left(\log\frac{\tau(s+1,r)}{\tau(s,r)}\right)_{t_1}\lambda^{-1}
    + f_2\lambda^{-2} + f_3\lambda^{-3} + \cdots\right),
\end{gather*}
where $f_n$'s denote the coef\/f\/icients of the expansion
\begin{gather*}
  \exp\left((e^{-D(\lambda)}-1)
     \log\frac{\tau(s+1,r)}{\tau(s,r)}\right)
   = \exp\left(\sum_{n=1}^\infty
     \lambda^{-n}h_n(-\tilde{\rd}_t)
     \log\frac{\tau(s+1,r)}{\tau(s,r)} \right)\\
\hphantom{\exp\left((e^{-D(\lambda)}-1)
     \log\frac{\tau(s+1,r)}{\tau(s,r)}\right)}{} = 1 + \sum_{n=1}^\infty f_n\lambda^{-n}.
\end{gather*}
Thus, expanding (\ref{lin-gen1}) in powers of $\lambda$,
we obtain the equation
\begin{gather}
- \rd_{t_1}\Phi_1(s,r,\mu)
+ \left(e^{\rd_s}
     + \left(\log\frac{\tau(s+1,r)}{\tau(s,r)}\right)_{t_1}
  \right)\Phi_1(s,r,\mu)\nonumber\\
\qquad- \frac{\tau(s+1,r+1)}{\tau(s,r)}e^{\rd_s}\Phi_2(s,r,\mu)
= 0
\label{lin-gen1-n=0}
\end{gather}
from the $\lambda^{-1}$ terms and the equations
\begin{gather}
h_{n+1}(-\tilde{\rd}_t)\Phi_1(s,r,\mu) - f_{n+1}\Phi_1(s,r,\mu)\nonumber\\
\qquad{}- \frac{\tau(s+1,r+1)}{\tau(s,r)}
         \sum_{m=0}^n f_mh_{n-m}(-\tilde{\rd}_t)
         e^{\rd_s}\Phi_2(s,r,\mu)
= 0
\label{lin-gen1-n>0}
\end{gather}
for $n = 1,2,\ldots$ from the $\lambda^{-n-1}$ terms.

In the same way, we can decompose
(\ref{lin-gen2}), (\ref{lin-gen3}) and (\ref{lin-gen4})
into and the equations
\begin{gather}
\rd_{t_1}\Phi_2(s,r,\mu)
+ \left(e^{-\rd_s}
      - \left(\log\frac{\tau(s-1,r)}{\tau(s,r)}\right)_{t_1}
  \right)\Phi_2(s,r,\mu)\nonumber\\
\qquad{}- \frac{\tau(s-1,r-1)}{\tau(s,r)}e^{-\rd_s}\Phi_1(s,r,\mu)
= 0,
\label{lin-gen2-n=0}
\\
- \rd_{\tbar_1}\Phi_1(s,r,\mu)
+ \frac{\tau(s+1,r)\tau(s-1,r)}{\tau(s,r)^2}e^{-\rd_s}\Phi_1(s,r,\mu)\nonumber\\
\qquad{}- \frac{\tau(s+1,r)\tau(s,r+1)}{\tau(s,r)^2}
         e^{\rd_s}\Phi_2(s,r,\mu)
= 0,
\label{lin-gen3-n=0}
\\
\rd_{\tbar_1}\Phi_2(s,r,\mu)
+ \frac{\tau(s-1,r)\tau(s+1,r)}{\tau(s,r)^2}
         e^{\rd_s}\Phi_2(s,r,\mu)\nonumber\\
\qquad{}- \frac{\tau(s-1,r)\tau(s,r-1)}{\tau(s,r)^2}e^{-\rd_s}\Phi_1(s,r,\mu)
= 0
\label{lin-gen4-n=0}
\end{gather}
and the equations
\begin{gather}
h_{n+1}(\tilde{\rd}_t)\Phi_2(s,r,\mu) - g_{n+1}\Phi_2(r,s,\mu)\nonumber\\
\qquad{}
- \frac{\tau(s-1,r-1)}{\tau(s,r)}
  \sum_{m=0}^n g_mh_{n-m}(\tilde{\rd}_t)e^{-\rd_s}\Phi_1(s,r,\mu)
= 0,
\label{lin-gen2-n>0}
\\
h_{n+1}(-\tilde{\rd}_{\tbar})\Phi_1(s,r,\mu)
  + \frac{\tau(s+1,r)\tau(s-1,r)}{\tau(s,r)^2}
    \fbar_ne^{-\rd_s}\Phi_1(s,r,\mu) \nonumber\\
\qquad{}
  - \frac{\tau(s+1,r)\tau (s,r+1)}{\tau(s,r)^2}
    \sum_{m=0}^n \fbar_mh_{n-m}(-\tilde{\rd}_{\tbar})
      e^{\rd_s}\Phi_2(s,r,\mu)
= 0,
\label{lin-gen3-n>0}
\\
h_{n+1}(\tilde{\rd}_{\tbar})\Phi_2(s,r,\mu)
+ \frac{\tau(s+1,r)\tau(s-1,r)}{\tau(s,r)^2}
  \gbar_ne^{\rd_s}\Phi_2(s,r,\mu)\nonumber\\
\qquad{}
- \frac{\tau(s-1,r)\tau(s,r-1)}{\tau(s,r)^2}
   \sum_{m=0}^n \gbar_mh_{n-m}(\tilde{\rd}_{\tbar})
     e^{-\rd_s}\Phi_1(s,r,\mu)
= 0
\label{lin-gen4-n>0}
\end{gather}
for $n = 1,2,\ldots$, where $g_n$, $\fbar_n$ and $\gbar_n$
are the coef\/f\/icients of the expansion
\begin{gather*}
  \exp\left((e^{D(\lambda)}-1)
     \log\frac{\tau(s-1,r)}{\tau(s,r)}\right)
   = 1 + \sum_{n=1}^\infty g_n\lambda^{-n},\\
  \exp\left((e^{-\Dbar(\lambda)}-1)
     \log\frac{\tau(s+1,r)}{\tau(s,r)}\right)
   = 1 + \sum_{n=1}^\infty \fbar_n\lambda^n,\\
  \exp\left((e^{\Dbar(\lambda)}-1)
     \log\frac{\tau(s-1,r)}{\tau(s,r)}\right)
   = 1 + \sum_{n=1}^\infty \gbar_n\lambda^n.
\end{gather*}

Among these equations,
the lowest ones (\ref{lin-gen1-n=0}), (\ref{lin-gen2-n=0}),
(\ref{lin-gen3-n=0}) and (\ref{lin-gen4-n=0})
can be cast into a matrix form, which agrees with
the lowest members (\ref{lin-Psi-t1}) and
(\ref{lin-Psi-tbar1}) of (\ref{lin-Psi-t}) and
(\ref{lin-Psi-tbar}). Though the other equations
(\ref{lin-gen1-n>0}), (\ref{lin-gen2-n>0}),
(\ref{lin-gen3-n>0}) and (\ref{lin-gen4-n>0})
do not take such an evolutionary form, they can be
recursively converted to the form of (\ref{lin-Psi-t})
just as in the case of the DKP hierarchy \cite{Takasaki07}.
Thus the auxiliary linear equations can be recovered
from the dif\/ference Fay identities.

\section{Dispersionless limit}\label{section5}

\subsection{Dispersionless Hirota equations}\label{section5.1}

As in the case of the KP and Toda hierarchies \cite{TT-review},
dispersionless limit is achieved by allowing
the tau function to depend on a small parameter
(Planck constant) $\hbar$ and assuming
the ``quasi-classical'' behavior
\begin{gather}
  \tau_\hbar(s,r,\bst,\bstbar)
  = e^{\hbar^{-2}F(s,r,\bst,\bstbar) + O(\hbar^{-1})}
  \qquad (\hbar \to 0)
\label{qc-tau}
\end{gather}
of the rescaled tau function
\begin{gather*}
  \tau_\hbar(s,r,\bst,\bstbar)
  = \tau(\hbar^{-1}s,\hbar^{-1}r,\hbar^{-1}\bst,\hbar^{-1}\bstbar).
\end{gather*}
As we show below, the dif\/ference Fay identities
for the rescaled tau function $\tau_\hbar(s,r,\bst,\bstbar)$
turn into dif\/ferential equations for the $F$-function
$F(s,r,\bst,\bstbar)$.  Following the terminology
commonly used in the literature, let us call
those equations ``dispersioness Hirota equations''.

Let us f\/irst consider (\ref{dFay1a}).
Upon multiplying both hand sides by
\begin{gather*}
  \frac{\tau(s+1,r+1,\bst,\bstbar)}
  {\tau(s+1,r+1,\bst+[\lambda^{-1}],\bstbar)
   \tau(s+1,r+1,\bst+[\mu^{-1}],\bstbar)
   \tau(s+1,r,\bst,\bstbar)},
\end{gather*}
this equation turns into such a form as
\begin{gather*}
\frac{\tau(s+1,r+1,\bst+[\lambda^{-1}]+[\mu^{-1}],\bstbar)
 \tau(s+1,r+1,\bst,\bstbar)}
{\tau(s+1,r+1,\bst+[\lambda^{-1}],\bstbar)
 \tau(s+1,r+1,\bst+[\mu^{-1}],\bstbar)}\\
= \frac{\lambda}{\lambda-\mu}
  \frac{\tau(s+1,r,\bst+[\lambda^{-1}],\bstbar)
   \tau(s+1,r+1,\bst,\bstbar)}
  {\tau(s+1,r+1,\bst+[\lambda^{-1}],\bstbar)
   \tau(s+1,r,\bst,\bstbar)}\\
 {}- \frac{\mu}{\lambda-\mu}
  \frac{\tau(s+1,r,\bst+[\mu^{-1}],\bstbar)
   \tau(s+1,r+1,\bst,\bstbar)}
  {\tau(s+1,r+1,\bst+[\mu^{-1}],\bstbar)
   \tau(s+1,r,\bst,\bstbar)}\\
 {}+ \frac{1}{\lambda\mu}
  \frac{\tau(s+2,r+1,\bst+[\lambda^{-1}]+[\mu^{-1}],\bstbar)
   \tau(s,r+1,\bst,\bstbar)}
  {\tau(s+1,r+1,\bst+[\lambda^{-1}],\bstbar)
   \tau(s+1,r+1,\bst+[\mu^{-1}],\bstbar)}
   \frac{\tau(s,r,\bst,\bstbar)\tau(s+1,r+1,\bst,\bstbar)}
        {\tau(s,r+1,\bst,\bstbar)\tau(s+1,r,\bst,\bstbar)}.
\end{gather*}
We can rewrite both hand sides as
\begin{gather*}
\mbox{LHS}
= \exp\big(\big(e^{D(\lambda)}-1\big)\big(e^{D(\mu)}-1\big)\log\tau(s+1,r+1)\big)
\end{gather*}
and
\begin{gather*}
\mbox{RHS}
=\frac{\lambda}{\lambda-\mu}
  \exp\big(\big(e^{-\rd_r}-1\big)\big(e^{D(\lambda)}-1\big)\log\tau(s+1,r+1)\big)\\
\phantom{\mbox{RHS}=}{}- \frac{\mu}{\lambda-\mu}
  \exp\big(\big(e^{-\rd_r}-1\big)\big(e^{D(\mu)}-1\big)\log\tau(s+1,r+1)\big)\\
\phantom{\mbox{RHS}=}{}+ \frac{1}{\lambda\mu}
  \exp\big(\big(e^{\rd_s+D(\lambda)}-1\big)\big(e^{\rd_s+D(\mu)}-1\big)e^{-\rd_s}
    \log\tau(s+1,r+1)\big)\\
\phantom{\mbox{RHS}=}{}\times
   \exp\big(\big(e^{-\rd_r}-1\big)\big(e^{-\rd_s}-1\big)\log\tau(s+1,r+1)\big).
\end{gather*}
Now rescale the variables $s$, $r$, $\bst$, $\bstbar$ as
\begin{gather}
  s \to \hbar^{-1}s,\qquad
  r \to \hbar^{-1}r,\qquad
  \bst \to \hbar^{-1}\bst,\qquad
  \bstbar \to \hbar^{-1}\bstbar.
\label{rescale-var}
\end{gather}
The last equation thereby becomes an equation
for the rescaled tau function $\tau_\hbar$,
in which the derivatives are rescaled as
\begin{gather}
  \rd_s \to \hbar\rd_s,\qquad
  \rd_r \to \hbar\rd_r,\qquad
  D(z) \to \hbar D(z),\qquad
  \Dbar(z) \to \hbar \Dbar(z).
\label{rescale-der}
\end{gather}
Under the quasi-classical ansatz (\ref{qc-tau}),
we can take the limit of this equation
as $\hbar \to 0$.  The outcome is the equation
\begin{gather}
e^{D(\lambda)D(\mu)F}
= \frac{\lambda e^{-\rd_rD(\lambda)F}
      - \mu e^{-\rd_rD(\mu)F}}{\lambda-\mu}
  + \frac{1}{\lambda\mu}
    e^{(\rd_s+D(\lambda)(\rd_s+D(\mu))F+\rd_r\rd_sF}.
\label{dHirota1a}
\end{gather}

In much the same way, we can derive
the following equations from the other
dif\/ferential Fay identities
(\ref{dFay1b})--(\ref{dFay3b}):
\begin{gather}
e^{D(\lambda)D(\mu)F}
= \frac{\lambda e^{-\rd_sD(\lambda)F}
      - \mu e^{-\rd_sD(\mu)F}}{\lambda-\mu}
  + \frac{1}{\lambda\mu}
    e^{(\rd_r+D(\lambda))(\rd_r+D(\mu))F+\rd_r\rd_sF},
\label{dHirota1b}
\\
e^{\Dbar(\lambda)\Dbar(\mu)F}
= \frac{\lambda^{-1}e^{-\rd_r\Dbar(\lambda)F}
      - \mu^{-1}e^{-\rd_r\Dbar(\mu)F}}{\lambda^{-1}-\mu^{-1}}
  + \lambda\mu
    e^{(-\rd_s+\Dbar(\lambda))(-\rd_s+\Dbar(\mu))F-\rd_r\rd_sF},
\label{dHirota2a}
\\
e^{\Dbar(\lambda)\Dbar(\mu)F}
= \frac{\lambda^{-1}e^{\rd_s\Dbar(\lambda)F}
      - \mu^{-1}e^{\rd_s\Dbar(\mu)F}}{\lambda^{-1}-\mu^{-1}}
  + \lambda\mu
    e^{(\rd_r+\Dbar(\lambda))(\rd_r+\Dbar(\mu))F-\rd_r\rd_sF},
\label{dHirota2b}
\\
e^{D(\lambda)\Dbar(\mu)F}
= e^{-(\rd_r+\rd_s)\Dbar(\mu)F}
  - \lambda\mu e^{-(\rd_rD(\lambda)+\rd_s\Dbar(\mu)+\rd_r\rd_s)F}\nonumber\\
  \phantom{e^{D(\lambda)\Dbar(\mu)F}=}{}+
    \lambda\mu e^{(-\rd_s+D(\lambda))(-\rd_s+\Dbar(\mu))F-(\rd_r+\rd_s)\rd_sF},
\label{dHirota3a}
\\
e^{D(\lambda)\Dbar(\mu)F}
= 1 - \frac{\mu}{\lambda}e^{\rd_s(\rd_s+D(\lambda)-\Dbar(\mu))F}
  \mbox{}+ \frac{\mu}{\lambda}
    e^{(\rd_r+D(\lambda))(\rd_r+\Dbar(\mu))F}.
\label{dHirota3b}
\end{gather}

The six dispersionless Hirota equations can be
divided to two distinct sets
$\{(\ref{dHirota1b}), (\ref{dHirota2b}),(\ref{dHirota3b})\}$
and $\{(\ref{dHirota1a}), (\ref{dHirota2a}),(\ref{dHirota3a})\}$.
The f\/irst set of equations may be thought of as analogues of
the dispersionless Hirota equations of the Toda hierarchy
\cite{Zabrodin01,KKMWWZ01,BMRWZ01,Teo03}.
The only dif\/ference is the presence of the last term
on the right hand side of each equation.
The second set of equations have no counterpart
in the Toda hierarchy.

We can use (\ref{dHirota1a}), (\ref{dHirota2a}) and
(\ref{dHirota3a}) to eliminate $e^{D(\lambda)D(\mu)F}$,
$e^{\Dbar(\lambda)\Dbar(\mu)F}$ and
$e^{D(\lambda)\Dbar(\mu)F}$ from (\ref{dHirota1b}),
(\ref{dHirota2b}) and (\ref{dHirota3b}).
This leads to the following remarkable observation.

\begin{theorem}
The dispersionless Hirota equations
\eqref{dHirota1a}--\eqref{dHirota3b}
imply the following equations:
\begin{gather}
\lambda\big(e^{-\rd_sD(\lambda)F} - e^{-\rd_rD(\lambda)F}\big)
+ \lambda^{-1}\big(e^{\rd_s(\rd_r+\rd_s+D(\lambda))F}
            - e^{\rd_r(\rd_r+\rd_s+D(\lambda))F}\big)\nonumber\\
 \qquad = (\rd_r - \rd_s)\rd_{t_1}F,
\label{dHirota1c}
\\
\lambda^{-1}\big(e^{\rd_s\Dbar(\lambda)F} - e^{-\rd_r\Dbar(\lambda)F}\big)
+ \lambda\big(e^{-\rd_s(\rd_r-\rd_s+\Dbar(\lambda))F}
        - e^{\rd_r(\rd_r-\rd_s+\Dbar(\lambda))F}\big)\nonumber\\
\qquad  = (\rd_r + \rd_s)\rd_{\tbar_1}F,
\label{dHirota2c}
\\
  (\rd_r + \rd_s)\rd_{\tbar_1}F
  = e^{-\rd_r\rd_sF}(\rd_r - \rd_s)\rd_{t_1}F.
\label{dHirota3c}
\end{gather}
\end{theorem}

\begin{proof}
We can rewrite (\ref{dHirota1a}) and (\ref{dHirota1b}) as
\begin{gather*}
  e^{D(\lambda)D(\mu)F}\left(1
    - \frac{1}{\lambda\mu}
      e^{\rd_s(\rd_r+\rd_s+D(\lambda)+D(\mu))F}\right)
  = \frac{\lambda e^{-\rd_rD(\lambda)F}
      - \mu e^{-\rd_rD(\lambda)F}}{\lambda-\mu},\\
  e^{D(\lambda)D(\mu)F}\left(1
    - \frac{1}{\lambda\mu}
      e^{\rd_r(\rd_r+\rd_s+D(\lambda)+D(\mu))F}\right)
  = \frac{\lambda e^{-\rd_sD(\lambda)F}
      - \mu e^{-\rd_sD(\mu)F}}{\lambda - \mu}.
\end{gather*}
Eliminating $e^{D(\lambda)D(\mu)F}$ from these equations
yields the equation
\begin{gather*}
\left(1 - \frac{1}{\lambda\mu}
          e^{\rd_s(\rd_r+\rd_s+D(\lambda)+D(\mu))F}\right)
\big(\lambda e^{-\rd_sD(\lambda)F} - \mu e^{-\rd_sD(\mu)F}\big)\\
\qquad= \left(1 - \frac{1}{\lambda\mu}
            e^{\rd_r(\rd_r+\rd_s+D(\lambda)+D(\mu))F}\right)
  \big(\lambda e^{-\rd_rD(\lambda)F} - \mu e^{-\rd_rD(\mu)F}\big),
\end{gather*}
which can be expanded as
\begin{gather*}
\lambda e^{-\rd_sD(\lambda)F} - \mu e^{-\rd_sD(\mu)F}
+ \lambda^{-1}e^{\rd_s(\rd_r+\rd_s+D(\lambda))F}
- \mu^{-1}e^{\rd_s(\rd_r+\rd_s+D(\mu))F}\\
\qquad{}= \lambda e^{-\rd_rD(\lambda)F} - \mu e^{-\rd_rD(\mu)F}
  + \lambda^{-1}e^{\rd_r(\rd_r+\rd_s+D(\lambda))F}
  - \mu^{-1}e^{\rd_r(\rd_r+\rd_s+D(\mu))F}.
\end{gather*}
We can separate $\lambda$-dependent and
$\mu$-dependent terms to each side
of the equation as
\begin{gather*}
\lambda\big(e^{-\rd_sD(\lambda)F} - e^{-\rd_rD(\lambda)F}\big)
+ \lambda^{-1}\big(e^{\rd_s(\rd_r+\rd_s+D(\lambda))F}
             - e^{\rd_r(\rd_r+\rd_s+D(\lambda))F}\big)\\
\qquad{}= \mu\big(e^{-\rd_sD(\mu)F} - e^{-\rd_rD(\mu)F}\big)
  + \mu^{-1}\big(e^{\rd_s(\rd_r+\rd_s+D(\mu))F}
             - e^{\rd_r(\rd_r+\rd_s+D(\mu))F}\big).
\end{gather*}
Therefore both hand sides of the last equation
are independent of $\lambda$ and $\mu$.
Letting $\lambda,\mu \to \infty$,
we can readily see that this quantity
is equal to $(\rd_r - \rd_s)\rd_{t_1}F$.
Thus we obtain (\ref{dHirota1c}).
In much the same way, we can derive (\ref{dHirota2c})
from (\ref{dHirota2a}) and (\ref{dHirota2b}).
If we start from (\ref{dHirota3a}) and (\ref{dHirota3b}),
we end up with the equation
\begin{gather*}
\mu^{-1}\big(e^{\rd_s\Dbar(\mu)F} - e^{-\rd_r\Dbar(\mu)F}\big)
+ \mu\big(e^{-\rd_s(\rd_r-\rd_s+\Dbar(\mu))F}
    - e^{\rd_r(\rd_r-\rd_s+\Dbar(\mu))F}\big)\\
\qquad{}= e^{-\rd_r\rd_sF}\big(
    \lambda\big(e^{-\rd_sD(\lambda)F} - e^{-\rd_rD(\lambda)F}\big)
      + \lambda^{-1}\big(e^{\rd_s(\rd_r+\rd_s+D(\lambda))F}
            - e^{\rd_r(\rd_r+\rd_s+D(\lambda))F}\big) \big).
\end{gather*}
By (\ref{dHirota1c}) and (\ref{dHirota2c}),
this equation reduces to (\ref{dHirota3c}).
\end{proof}

It is easy to see from this proof that the converse
is also true.  Namely, if (\ref{dHirota1b}), (\ref{dHirota2b})
and (\ref{dHirota3b}) holds, the other three
dispersionless Hirota equations (\ref{dHirota1a}),
(\ref{dHirota2a}) and (\ref{dHirota3a})
can be recovered from (\ref{dHirota1c}),
(\ref{dHirota2c}) and (\ref{dHirota3c}).
Thus we are led to the following conclusion.

\begin{corollary}
The dispersionless Hirota equations
\eqref{dHirota1a}--\eqref{dHirota3b},
can be reduced to the coupled system
of the two sets of equations
$\{(\ref{dHirota1b}),(\ref{dHirota2b}), (\ref{dHirota3b})\}$
and $\{(\ref{dHirota1c}),(\ref{dHirota2c}),(\ref{dHirota3c})\}$.
\end{corollary}

Thus the dispersionless Hirota equations
of the Pfaf\/f--Toda hierarchy can be reduced to
two sets of equations that have quite dif\/ferent
appearance and nature.  Among the second set
of equations, the last equation (\ref{dHirota3c})
is nothing but the dispersionless limit of
(\ref{Hirota-lowest2}).  The other two equations
(\ref{dHirota1c}) and (\ref{dHirota2c}) appear
to be more mysterious.  As it turns out below,
they are related to special auxiliary
linear equations for the wave functions.

\subsection{Elliptic spectral curve}\label{section5.2}

Let us examine (\ref{dHirota1c}), (\ref{dHirota2c})
and (\ref{dHirota3c}) in more detail.
Substituting $\lambda \to z$, and applying (\ref{dHirota3c})
to the term on the right hand side of (\ref{dHirota2c}),
we can rewrite (\ref{dHirota1c}) and (\ref{dHirota2c}) as
\begin{gather*}
z\big(e^{-\rd_sD(z)F} - e^{-\rd_rD(z)F}\big)
+ z^{-1}\big(e^{\rd_s(\rd_r+\rd_s+D(z))F}
            - e^{\rd_r(\rd_r+\rd_s+D(z))F}\big)
  = (\rd_r - \rd_s)\rd_{t_1}F,
\\
z^{-1}\big(e^{\rd_s\Dbar(z)F} \!- e^{-\rd_r\Dbar(z)F}\big)\!
+ z\big(e^{-\rd_s(\rd_r{-}\rd_s{+}\Dbar(z))F}\!
        - e^{\rd_r(\rd_r{-}\rd_s{+}\Dbar(z))F}\big)
 \! = e^{-\rd_r\rd_sF}\!(\rd_r\! - \rd_s)\rd_{t_1}\!F.\!
\end{gather*}
Actually, these two equations are equivalent to
the set of three equations (\ref{dHirota1c}),
(\ref{dHirota2c}) and~(\ref{dHirota3c}),
because letting $z \to 0$ in the second equation
yields~(\ref{dHirota3c}), and this enables one
to recover (\ref{dHirota2c}) as well.

A clue of the subsequent consideration
is to introduce the auxiliary functions
\begin{gather*}
  S(z) = \xi(\bst,z) + (s+r)\log z - D(z)F,\\
  \Sbar(z) = \xi(\bstbar,z^{-1}) + (s-r)\log z
              + \rd_sF - \Dbar(z)F.
\end{gather*}
We can thereby rewrite these equations into
the ``Hamilton--Jacobi'' form
\begin{gather}
  e^{\rd_rS(z)}
  = e^{\rd_sS(z)} + (\rd_s-\rd_r)\rd_{t_1}F
  + e^{\rd_s(\rd_r+\rd_s)F}e^{-\rd_sS(z)}
  - e^{\rd_r(\rd_r+\rd_s)F}e^{-\rd_rS(z)},
\label{HJ-S-r}\\
  e^{\rd_r\Sbar(z)}
  = e^{\rd_s\Sbar(z)} + (\rd_s-\rd_r)\rd_{t_1}F
  + e^{\rd_s(\rd_r+\rd_s)F}e^{-\rd_s\Sbar(z)}
  - e^{\rd_r(\rd_r+\rd_s)F}e^{-\rd_r\Sbar(z)}.
\label{HJ-Sbar-r}
\end{gather}
These equations imply that the exponentiated
gradient vectors $(e^{\rd_sS(z)},e^{\rd_rS(z)})$
and $(e^{\rd_s\Sbar(z)}$, $e^{\rd_r\Sbar(z)})$ of
the two $S$-functions on the $(s,r)$ plane
satisfy the same algebraic equation
\begin{gather}
  Q = P + (\rd_s-\rd_r)\rd_{t_1}F
  + e^{\rd_s(\rd_r+\rd_s)F}P^{-1}
  - e^{\rd_r(\rd_r+\rd_s)F}Q^{-1}.
\label{spec-curve}
\end{gather}
This equation def\/ines a curve of genus one
on the $(P,Q)$ plane.

We now argue that this equation can be identif\/ied
with the characteristic equation of
a set of auxiliary linear equations.
In other words, this curve is indeed
a ``spectral curve''.

The functions $S(z)$ and $\Sbar(z)$
may be thought of as the phase functions
of the quasi-classical (WKB) ansatz
\begin{gather*}
  \Psi_1(z) = e^{\hbar^{-1}S(z) + O(1)}, \qquad
  \Psibar_1(z) = e^{\hbar^{-1}\Sbar(z) + O(1)} \qquad
  (\hbar \to 0)
\end{gather*}
of the wave functions $\Psi_1(z)$ and $\Psibar_1(z)$.
Consequently, it will be natural to expect
that (\ref{HJ-S-r}) and (\ref{HJ-Sbar-r})
are Hamilton--Jacobi equations of linear equations
of the form
\begin{gather}
  e^{\hbar\rd_r}\Phi_1(z)
  = \big(e^{\hbar\rd_s} + a + be^{-\hbar\rd_s}
    + ce^{-\hbar\rd_r}\big)\Phi_1(z)
\label{lin-scalar-r}
\end{gather}
for $\Phi_1(z) = \Psi_1(z),\Psibar_1(z)$,
and that the coef\/f\/icients have the quasi-classical limit
\begin{gather*}
  \lim_{\hbar\to 0}a = (\rd_s-\rd_r)F,\qquad
  \lim_{\hbar\to 0}b = e^{\rd_s(\rd_r+\rd_s)F},\qquad
  \lim_{\hbar\to 0}c = - e^{\rd_r(\rd_r+\rd_s)F}
\end{gather*}
so that (\ref{spec-curve}) can be interpreted
as the characteristic equation of~(\ref{lin-scalar-r}).

We can indeed derive such linear equations
from (\ref{lin-Psi-r}).  To simplify notations,
let us consider the case where $\hbar = 1$;
one can readily transfer to the $\hbar$-dependent setup of
the quasi-classical ansatz by rescaling the variables
and derivative as (\ref{rescale-var}) and (\ref{rescale-der}).
Written in terms of components, (\ref{lin-Psi-r})
consists of the linear equations
\begin{gather*}
e^{\rd_r}\Phi_1(z) = A\Phi_1(z) + B\Phi_2(z),\qquad
e^{\rd_r}\Phi_2(z) = C\Phi_1(z)
\end{gather*}
for the four pairs $\Phi_\alpha(z) = \Psi_\alpha(z),
\Psi^*_\alpha(z),\Psibar_\alpha(z),\Psibar^*_\alpha(z)$
($\alpha = 1,2$), where
\begin{gather*}
  A = e^{\rd_s}
    + \left(\log\frac{\tau(s+1,r)}{\tau(s,r+1)}\right)_{t_1}
    + \frac{\tau(s+1,r+1)\tau(s-1,r)}{\tau(s,r+1)\tau(s,r)}e^{-\rd_s},\nonumber\\
  B = - \frac{\tau(s+1,r+1)}{\tau(s,r)}e^{\rd_s}, \qquad
  C = \frac{\tau(s-1,r)}{\tau(s,r+1)}e^{-\rd_s}.
\end{gather*}
The second equation implies that $\Phi_1(z)$
and $\Phi_2(z)$ are connected by the relation
\begin{gather*}
  \Phi_2(z)
  = e^{-\rd_r}C\Phi_1(z)
  = \frac{\tau(s-1,r-1)}{\tau(s,r)}e^{-\rd_s-\rd_r}\Phi_1(z),
\end{gather*}
which one can see from the def\/inition
(\ref{wave-fn1}) and (\ref{wave-fn2})
of the wave functions as well.
Using this relation, we can eliminate $\Phi_2(z)$
from the f\/irst equation and obtain the equation
\begin{gather*}
  e^{\rd_r}\Phi_1(z) = (A + Be^{-\rd_r}C)\Phi_1(z).
\end{gather*}
After some algebra, this equation boils down
to (\ref{lin-scalar-r}).  Moreover, though we omit details,
the coef\/f\/icients turn out to take such a form as
\begin{gather*}
  a = \left(\log\frac{\tau(s+1,r)}{\tau(s,r+1)}\right)_{t_1},\qquad
  b = \frac{\tau(s+1,r+1)\tau(s-1,r)}{\tau(s,r+1)\tau(s,r)},\\
  c  = - \frac{\tau(s+1,r+1)\tau(s,r-1)}{\tau(s+1,r)\tau(s,r)}.
\end{gather*}
It is easy to see that these coef\/f\/icients
do have the anticipated quasi-classical limit.

The other auxiliary linear equations
(\ref{lin-Psi-t}) and (\ref{lin-Psi-tbar}), too,
can be converted to scalar linear equations of the form
\begin{gather*}
  \rd_{t_n}\Phi_1(z) = (A_n + B_ne^{-\rd_r}C)\Phi_1(z), \qquad
  \rd_{\tbar_n}\Phi_1(z) = (\Abar_n + \Bbar_ne^{-\rd_r}C)\Phi_1(z).
\end{gather*}
Note that the $2D$ dif\/ference operators
\begin{gather*}
  K_n(e^{\rd_s},e^{\rd_r}) = A_n + B_ne^{-\rd_r}C, \qquad
  \Kbar_n(e^{\rd_s},e^{\rd_r}) = \Abar_n + \Bbar_ne^{-\rd_r}C
\end{gather*}
show up on the right hand sides of these equations.
Consequently, the S-functions~$S(z)$ and~$\Sbar(z)$
satisfy the Hamilton--Jacobi equations
\begin{gather}
  \rd_{t_n}S(z) = K_n\big(e^{\rd_sS(z)},e^{\rd_rS(z)}\big),\qquad
  \rd_{\tbar_n}S(z) = \Kbar_n\big(e^{\rd_sS(z)},e^{\rd_rS(z)}\big),
\label{HJ-S-ttbar}
\\
  \rd_{t_n}\Sbar(z) = K_n\big(e^{\rd_s\Sbar(z)},e^{\rd_r\Sbar(z)}\big),\qquad
  \rd_{\tbar_n}\Sbar(z) = \Kbar_n\big(e^{\rd_s\Sbar(z)},e^{\rd_r\Sbar(z)}\big).
\label{HJ-Sbar-ttbar}
\end{gather}
Those Hamilton--Jacobi equations and
the constraints (\ref{HJ-S-r}) and (\ref{HJ-Sbar-r})
may be thought of as def\/ining ``quasi-classical deformations''
\cite{KKMA04,KKMA05} of the spectral curve (\ref{spec-curve}).

\subsection{Comparison with DKP hierarchy}\label{section5.3}

Let us compare the foregoing results with the case of
the DKP hierarchy \cite{Takasaki07,KP08}.

The situation of the DKP hierarchy is similar
to the KP hierarchy rather than the Toda hierarchy.
The role of the dif\/ference Fay identities
(\ref{dFay1a})--(\ref{dFay3b}) are played
by the dif\/ferential Fay identities
\begin{gather}
  \frac{\tau(r,\bst+[\lambda^{-1}]+[\mu^{-1}])\tau(r,\bst)}
       {\tau(r,\bst+[\lambda^{-1}])\tau(r,\bst+[\mu^{-1}])}
  - \frac{1}{\lambda^2\mu^2}
    \frac{\tau(r+1,\bst+[\lambda^{-1}]+[\mu^{-1}])\tau(r-1,\bst)}
         {\tau(r,\bst+[\lambda^{-1}])\tau(r,\bst+[\mu^{-1}])}\nonumber \\
\qquad{}  = 1
    - \frac{1}{\lambda-\mu}\rd_{t_1}\log
      \frac{\tau(r,\bst+[\lambda^{-1}])}
           {\tau(r,\bst+[\mu^{-1}])},
\label{DKP-dFay1}
\\
  \frac{\lambda^2}{\lambda-\mu}
  \frac{\tau(r-1,\bst+[\lambda^{-1}])\tau(r,\bst+[\mu^{-1}])}
       {\tau(r,\bst+[\lambda^{-1}]+[\mu^{-1}])\tau(r-1,\bst)}
- \frac{\mu^2}{\lambda-\mu}
  \frac{\tau(r,\bst+[\lambda^{-1}])\tau(r-1,\bst+[\mu^{-1}])}
       {\tau(r,\bst+[\lambda^{-1}]+[\mu^{-1}])\tau(r-1,\bst)}\nonumber \\
\qquad= \lambda + \mu
  - \rd_{t_1}\log
    \frac{\tau(r,\bst+[\lambda^{-1}]+[\mu^{-1}])}{\tau(r-1,\bst)}.
\label{DKP-dFay2}
\end{gather}
In the dispersionless limit, they turn
into the dispersionless Hirota equations
\begin{gather}
  e^{D(\lambda)D(\mu)F}
  - \lambda^{-2}\mu^{-2}
    e^{(\rd_r+D(\lambda))(\rd_r+D(\mu)F}
  = 1
    - \frac{\rd_{t_1}(D(\lambda)-D(\mu))F}{\lambda-\mu},
\label{DKP-dHirota1}
\\
  \frac{\lambda^2}{\lambda-\mu}e^{-D(\lambda)(\rd_r+D(\mu))F}\!
  - \frac{\mu^2}{\lambda-\mu}e^{-(\rd_r+D(\lambda))D(\mu)F}
  = \lambda + \mu
    - \rd_{t_1}(\rd_r+D(\lambda)+D(\mu))F\!\!\!
\label{DKP-dHirota2}
\end{gather}
for the $F$-function $F(r,\bst)$.
The f\/irst equation (\ref{DKP-dHirota1})
resembles the dispersionless Hirota equation
of the KP hierarchy.

It is convenient to introduce the $S$-function
\begin{gather*}
  S(z) = \xi(\bst,z) + 2r\log z - D(z)F
\end{gather*}
at this state.  We can thereby rewrite
(\ref{DKP-dHirota1}) and (\ref{DKP-dHirota2}) as
\begin{gather*}
  e^{D(\lambda)D(\mu)F}
  \big(1 - e^{\rd_r^2F-\rd_rS(\lambda)-\rd_rS(\mu)}\big)
  = \frac{\rd_1S(\lambda)-\rd_1S(\mu)}{\lambda-\mu},
\\
  \frac{e^{-D(\lambda)D(\mu)F}}{\lambda-\mu}
  \big(e^{\rd_rS(\lambda)}-e^{\rd_rS(\mu)}\big)
  = \rd_{t_1}S(\lambda) + \rd_{t_1}S(\mu) - \rd_r\rd_{t_1}F
\end{gather*}
and eliminate $e^{D(\lambda)D(\mu)F}$ from
these two equations.  This leads to the equality
\begin{gather*}
  (\rd_{t_1}S(\lambda))^2 - (\rd_{t_1}\rd_rF)(\rd_{t_1}S(\lambda))
  - e^{\rd_rS(\lambda)} - e^{\rd_r^2F-\rd_rS(\lambda)} \\
\qquad{} = (\rd_{t_1}S(\mu))^2 - (\rd_{t_1}\rd_rF)(\rd_{t_1}S(\mu))
  - e^{\rd_rS(\mu)} - e^{\rd_r^2F-\rd_rS(\mu)},
\end{gather*}
hence both hand sides are independent of $\lambda$
and $\mu$.  Letting $\lambda,\mu \to \infty$,
we can determine this quantity explicitly.
Thus we obtain the equation
\begin{gather}
  (\rd_{t_1}S(z))^2 - (\rd_{t_1}\rd_rF)(\rd_{t_1}S(z))
  - e^{\rd_rS(z)} - e^{\rd_r^2F-\rd_rS(z)} \nonumber\\
 \qquad{} = - 2\rd_{t_1}^2F + \frac{1}{2}\rd_{t_2}\rd_rF
    - \frac{1}{2}(\rd_{t_1}\rd_rF)^2.
\label{DKP-dHirota3}
\end{gather}
In other words, the partially exponentiated
gradient vector $(\rd_{t_1}S(z),e^{\rd_rS(z)})$
of the $S$-function satisf\/ies the algebraic equation
\begin{gather}
  p^2 - (\rd_{t_1}\rd_rF)p - Q - e^{\rd_r^2F}Q^{-1}
  = - 2\rd_{t_1}^2F + \frac{1}{2}\rd_{t_2}\rd_rF
    - \frac{1}{2}(\rd_{t_1}\rd_rF)^2
\label{DKP-spec-curve}
\end{gather}
of an elliptic curve on the $(p,Q)$ plane.
This curve was studied by Kodama and Pierce \cite{KP08}
as an analogue of the spectral curve of the $1D$ Toda hierarchy.

(\ref{DKP-dHirota3}) and (\ref{DKP-spec-curve})
are the Hamilton--Jacobi and characteristic equations
of a linear equation of the form
\begin{gather}
  e^{\hbar\rd_r}\Phi_1(z)
  = \big(\hbar^2\rd_{t_1}^2 + \hbar a\rd_{t_1} + b
     + ce^{-\hbar\rd_r}\big)\Phi_1(z).
\label{DKP-lin-scalar-r}
\end{gather}
We can derive this equation from one of auxiliary
linear equations  as follows.  Let us again consider
the case where $\hbar = 1$.  As in the case of
the Pfaf\/f--Toda hierarchy, we can eliminate
$\Psi_2(z)$ and $\Psi^*_2(z)$ from the auxiliary
linear equation (\ref{DKP-lin-Psi-r}) by the relation
\begin{gather*}
  \Phi_2(z) = \frac{\tau(r-1)}{\tau(r)}e^{-\rd_r}\Phi_1(z)
\end{gather*}
that holds for $\Phi_\alpha (z) = \Psi_\alpha(z),\Psi^*_\alpha(z)$
($\alpha = 1,2$).  The matrix equation (\ref{DKP-lin-Psi-r})
thereby reduces to a scalar equation of the form
(\ref{DKP-lin-scalar-r}) for $\Phi_1(z) = \Psi_1(z),\Psi^*_1(z)$.
The coef\/f\/icients $a$, $b$, $c$ can be determined explicitly as
\begin{gather}
  a = - \left(\log\frac{\tau(r+1)}{\tau(r)}\right)_{t_1},\qquad
  c = - \frac{\tau(r+1)\tau(r-1)}{\tau(r)^2},\nonumber\\
  b = \frac{\tau(r+1)_{t_1t_1} - \tau(r+1)_{t_2}}{2\tau(r+1)}
          - \frac{\tau(r)_{t_1t_1} - \tau(r)_{t_2}}{2\tau(r)}\nonumber\\
   \phantom{b=}{}- \left(\log\frac{\tau(r+1)}{\tau(r)}\right)_{t_1}
            (\log\tau(r))_{t_1}
          + 2(\log\tau(r))_{t_1t_1} .
\label{DKP-abc}
\end{gather}
Rescaling the variables as (\ref{rescale-var}),
one can correctly recover the coef\/f\/icients of
(\ref{DKP-dHirota3}) and (\ref{DKP-spec-curve})
in the quasi-classical limit:
\begin{gather*}
  \lim_{\hbar\to 0}a = - \rd_{t_1}\rd_rF, \qquad
  \lim_{\hbar\to 0}c = - e^{\rd_r^2F}, \qquad
  \lim_{\hbar\to 0}b
  = - \frac{1}{2}\rd_{t_2}\rd_rF
    + \frac{1}{2}(\rd_{t_1}\rd_rF)^2 + 2\rd_{t_1}^2F .
\end{gather*}

\section{Conclusion}\label{section6}

We have thus obtained the following equations
that characterize various aspects of
the Pfaf\/f--Toda hierarchy:
\begin{itemize}\itemsep=0pt
\item the algebraic constraints
(\ref{WV-constraint1}), (\ref{WV-constraint2})
and the evolution equations (\ref{WV-evolve-r}),
(\ref{WV-evolve-t}), (\ref{WV-evolve-tbar})
of the dressing operators,
\item the auxiliary linear equations
(\ref{lin-Psi-r})--(\ref{lin-Psi-tbar}),
\item the dif\/ference Fay identities
(\ref{dFay1a})--(\ref{dFay3b}),
\item the generating functional expression
(\ref{lin-gen1})--(\ref{lin-gen4})
of the auxiliary linear equations
(\ref{lin-Psi-t}) and (\ref{lin-Psi-tbar}),
\item the dispersionless Hirota equations,
(\ref{dHirota1a})--(\ref{dHirota3b}),
\item the def\/ining equation (\ref{spec-curve})
of the elliptic spectral curve,
\item the Hamilton--Jacobi equations
(\ref{HJ-S-r}), (\ref{HJ-Sbar-r}), (\ref{HJ-S-ttbar})
and (\ref{HJ-Sbar-ttbar}).
\end{itemize}
All these equations have counterparts in the DKP hierarchy.
We have thus demonstrated that the Pfaf\/f--Toda hierarchy
is indeed a Toda version of the DKP hierarchy
(or a Pfaf\/f\/ian version of the Toda hierarchy).

Actually, this is not the end of the story.
Let us note a few open problems.

Firstly,  although the other auxiliary linear equations
(\ref{lin-Psi-t}) and (\ref{lin-Psi-tbar}) have been
encoded to the dif\/ference Fay identities,
the status of the remaining equation (\ref{lin-Psi-r})
is still obscure.  Since its counterpart in the dispersionless limit
are (\ref{HJ-S-r}) and (\ref{HJ-Sbar-r}), and these equations
are obtained from the dispersionless Hirota equations,
it seems likely that (\ref{lin-Psi-r}), too, can be derived
from the dif\/ference Fay identities.  Unfortunately,
we have been unable to f\/ind a direct proof.
If this conjecture is true, it leads to an important conclusion
that the dif\/ference Fay identities are, on the whole,
equivalent to the Pfaf\/f--Toda hierarchy itself,
as it is indeed the case for the KP hierarchy
\cite{Takasaki07,TT-review} and the Toda hierarchy \cite{Teo06}.

Secondly, very little is known about special solutions
of the Pfaf\/f--Toda hierarchy.  Of course one can freely
generate solutions by the fermionic formula.
Finding an interesting class of solutions is, however,
a nontrivial problem.   A possible strategy will be
to seek, again, for analogy with the DKP hierarchy.

\appendix

\section{Auxiliary linear problem of DKP hierarchy}\label{appendixA}

The DKP hierarchy has a discrete variable $r$
and a set of continuous variables $\bst = (t_1,t_2,\ldots)$.
The Tau function $\tau = \tau(r,\bst)$ satisf\/ies
the bilinear equation
\begin{gather}
\oint\frac{dz}{2\pi i}
  z^{2r'-2r}e^{\xi(\bst'-\bst,z)}
  \tau(r',\bst'-[z^{-1}])\tau(r,\bst+[z^{-1}]) \nonumber\\
\qquad{}
+ \oint\frac{dz}{2\pi i}
    z^{2r-2r'-4}e^{\xi(\bst-\bst',z)}
    \tau(r'+1,\bst'+[z^{-1}])\tau(r-1,\bst-[z^{-1}])
= 0.
\label{DKP-bilin-tau}
\end{gather}
This bilinear equation can be converted to
the Hirota form
\begin{gather*}
  \sum_{n=0}^\infty h_n(-2\bsa)
  h_{n+2r'-2r+1}(\tilde{D}_{\bst})e^{\langle\bsa,D_{\bst}\rangle}
  \tau(r,\bst)\cdot\tau(r',\bst) \\
\qquad{}
+ \sum_{n=0}^\infty h_n(2\bsa)
  h_{n+2r-2r'-3}(-\tilde{D}_{\bst})e^{\langle\bsa,D_{\bst}\rangle}
  \tau(r-1,\bst)\cdot\tau(r'+1,\bst)
= 0
\end{gather*}
with an inf\/inite set of arbitrary constants
$\bsa = (a_1,a_2,\ldots)$.
This is a generating functional expression
of an inf\/inite number of Hirota equations.
The dif\/ferential Fay identities~(\ref{DKP-dFay1}) and~(\ref{DKP-dFay2})
can be derived by dif\/ferentiating
the bilinear equation by $t'_1$
and specializing the variables as follows:
\begin{itemize}\itemsep=0pt
\item[1)] $\bst' = \bst + [\lambda^{-1}] + [\mu^{-1}]$, $r' = r$;
\item[2)] $\bst' = \bst + [\lambda^{-1}] - [\mu^{-1}]$, $r' = r-1$.
\end{itemize}

The bilinear equation (\ref{DKP-bilin-tau})
of the tau function leads to a set of
bilinear equations for the wave functions
\begin{gather*}
  \Psi_1(r,\bst,z)  = z^{2r}e^{\xi(\bst,z)}
     \frac{\tau(r,\bst-[z^{-1}])}{\tau(r,\bst)},\nonumber\\
  \Psi_2(r,\bst,z)  = z^{2r-2}e^{\xi(\bst,z)}
     \frac{\tau(r-1,\bst-[z^{-1}])}{\tau(r,\bst)},\nonumber\\
  \Psi^*_1(r,\bst,z)  = z^{-2r-2}e^{-\xi(\bst,z)}
     \frac{\tau(r+1,\bst+[z^{-1}])}{\tau(r,\bst)},\nonumber\\
  \Psi^*_2(r,\bst,z)  = z^{-2r}e^{-\xi(\bst,z)}
     \frac{\tau(r,\bst+[z^{-1}])}{\tau(r,\bst)}.
\end{gather*}
These bilinear equations can be cast into
the matrix form
\begin{gather}
  \oint\frac{dz}{2\pi i}
  \Psi_{2\times 2}(r',\bst',z)\tp{\Psi^*_{2\times 2}(r,\bst,z)}
  = 0,
\label{DKP-bilin-Psi}
\end{gather}
where
\begin{gather*}
\Psi_{2\times 2}(r,\bst,z)
 = \left(\!\begin{array}{cc}
   \Psi_1(r,\bst,z) & \Psi^*_1(r,\bst,z)\\
   \Psi_2(r,\bst,z) & \Psi^*_2(r,\bst,z)
   \end{array}\!\right),\qquad
\Psi^*_{2\times 2}(r,\bst,z)
 = \left(\!\begin{array}{cc}
   \Psi^*_1(r,\bst,z) & \Psi_1(r,\bst,z)\\
   \Psi^*_2(r,\bst,z) & \Psi_2(r,\bst,z)
   \end{array}\!\right).
\end{gather*}

The wave functions are associated with
dressing operators of the form
\begin{gather*}
  W_1 = 1 + \sum_{n=1}^\infty w_{1n}\rd_{t_1}^{-n}, \qquad
  V_1 = \sum_{n=0}^\infty v_{1n}(-\rd_{t_1})^{-n-2},\\
  W_2 = \sum_{n=0}^\infty w_{2n}\rd_{t_1}^{-n-2}, \qquad
  V_2 = 1 + \sum_{n=1}^\infty v_{2n}(-\rd_{t_1})^{-n}.
\end{gather*}
The coef\/f\/icients are determined by Laurent expansion
of the tau-quotient in the wave functions as
\begin{gather*}
  \frac{\tau(r,\bst-[z^{-1}])}{\tau(r,\bst)}
    = 1 + \sum_{n=1}^\infty w_{1n}z^{-n},\qquad
  \frac{\tau(r+1,\bst+[z^{-1}])}{\tau(r,\bst)}
    = \sum_{n=0}^\infty v_{1n}z^{-n},\\
  \frac{\tau(r-1,\bst-[z^{-1}])}{\tau(r,\bst)}
    = \sum_{n=0}^\infty w_{2n}z^{-n},\qquad
  \frac{\tau(r,\bst+[z^{-1}])}{\tau(r,\bst)}
    = 1 + \sum_{n=1}^\infty v_{2n}z^{-n}.
\end{gather*}
The wave functions are thereby expressed as
\begin{gather*}
  \Psi_\alpha(r,\bst,z) = W_\alpha z^{2r}e^{\xi(\bst,z)},\qquad
  \Psi^*_\alpha(r,\bst,z) = V_\alpha z^{-2r}e^{-\xi(\bst,z)}
\end{gather*}
for $\alpha = 1,2$.

Various equations for the dressing operators
can be derived from this bilinear equation.
A~technical clue is an analogue of (\ref{oint-formula})
for pseudo-dif\/ferential operators \cite{DJKM-review,Dickey-book,KvdL93}.
For a pair of pseudo-dif\/ferential operators of the form
\begin{gather*}
  P = \sum_{n=-\infty}^\infty p_n(x)\rd_x^n,\qquad
  Q = \sum_{n=-\infty}^\infty q_n(x)\rd_x^n,
\end{gather*}
let $\Psi(x,z)$ and $\Phi(x,z)$ denote
the wave functions
\begin{gather*}
  \Psi(x,z) = Pe^{xz}
    = \sum_{n=-\infty}^\infty p_n(x)z^ne^{xz},\qquad
  \Phi(x,z) = Qe^{-xz}
    = \sum_{n=-\infty}^\infty q_n(x)(-z)^ne^{-xz}.
\end{gather*}
Moreover, let $P^*$ denote
the formal adjoint
\begin{gather*}
  P^* = \sum_{n=-\infty}^\infty (-\rd_x)^np_n(x).
\end{gather*}
Then one has the identity
\begin{gather}
  \oint\frac{dz}{2\pi i}\Psi(x',z)\Phi(x,z)
  = \sum_{k=0}^\infty (PQ^*)_{-k-1}\frac{(x'-x)^k}{k!}
 = - \sum_{k=0}^\infty (QP^*)_{-k-1}\frac{(x-x')^k}{k!},
\label{DKP-oint-formula}
\end{gather}
where $( \ )_{-k-1}$ stands for the coef\/f\/icient
of $\rd_x^{-k-1}$ of a pseudo-dif\/ferential operator.

With the aid of this formula (\ref{DKP-oint-formula}),
one can derive the algebraic constraint
\begin{gather}
  \left(\begin{array}{cc}
  W_1 & V_1 \\
  W_2 & V_2
  \end{array}\right)^*
= \left(\begin{array}{cc}
  0 & 1 \\
  -1 & 0
  \end{array}\right)
  \left(\begin{array}{cc}
  W_1 & V_1\\
  W_2 & V_2
  \end{array}\right)^{-1}
  \left(\begin{array}{cc}
  0 & -1 \\
  1 & 0
  \end{array}\right),
\label{DKP-W-constraint}
\end{gather}
the discrete evolution equation
\begin{gather}
  \left(\begin{array}{cc}
  W_1(r+1) & V_1(r+1)\\
  W_2(r+1) & V_2(r+1)
  \end{array}\right)
  \left(\begin{array}{cc}
  \rd_{t_1}^2 & 0\\
  0 & \rd_{t_1}^{-2}
  \end{array}\right)
= \left(\begin{array}{cc}
  A & B\\
  C & 0
  \end{array}\right)
  \left(\begin{array}{cc}
  W_1 & V_1\\
  W_2 & V_2
  \end{array}\right)
\label{DKP-W-evolve-r}
\end{gather}
and the continuous evolution equations
\begin{gather}
  \left(\begin{array}{cc}
  W_{1,t_n} + W_1\rd_{t_1} & V_{1,t_n} - V_2(-\rd_{t_1})^n\\
  W_{2,t_n} + W_2\rd_{t_1} & V_{2,t_n} - V_2(-\rd_{t_1})^n
  \end{array}\right)
= \left(\begin{array}{cc}
  A_n & B_n\\
  C_n & D_n
  \end{array}\right)
  \left(\begin{array}{cc}
  W_1 & V_1\\
  W_2 & V_2
  \end{array}\right)
\label{DKP-W-evolve-t}
\end{gather}
from the bilinear equation (\ref{DKP-bilin-Psi})
of the wave functions.
Here $A$, $B$, $C$ are dif\/ferential operators of the form
\begin{gather*}
  A = \rd_{t_1}^2 + a\rd_{t_1} + b,\qquad
  B = - \frac{\tau(r+1)}{\tau(r)},\qquad
  C = \frac{\tau(r)}{\tau(r+1)},
\end{gather*}
where $a$ and $b$ are the same quantities
as shown in (\ref{DKP-abc}).
$A_n$, $B_n$, $C_n$ are given by
\begin{gather*}
  A_n = (W_1\rd_{t_1}^nV_2^* + V_1(-\rd_{t_1})^nW_2^*)_{\ge 0}, \qquad
  B_n = - (W_1\rd_{t_1}^nV_1^* + V_1(-\rd_{t_1})^nW_1^*)_{\ge 0},\\
  C_n = (W_2\rd_{t_1}^nV_2^* + V_2(-\rd_{t_1})^nW_2^*)_{\ge 0}, \qquad
  D_n = - (W_2\rd_{t_1}^nV_1^* + V_2(-\rd_{t_1})^nW_1^*)_{\ge 0},
\end{gather*}
where $( \ )_{\ge 0}$ stands for the projection
onto nonnegative powers of $\rd_{t_1}$.
These operators satisfy the algebraic relations
\begin{gather*}
  A_n^* = - D_n, \qquad  B_n^* = B_n,\qquad
  C_n^* = C_n, \qquad D_n^* = - A_n,
\end{gather*}
which may be thought of as Lie algebraic counterparts
of the constraint (\ref{DKP-W-constraint})
for the dressing operators.  This algebraic structure
is generalized by Kac and van de Leur \cite{KvdL98}
to multicomponent hierarchies.  Let us mention that
these algebraic relations among $A_n$, $B_n$, $C_n$, $D_n$
are also derived by Kakei \cite{Kakei00}
in a inverse scattering formalism.

The evolution equations (\ref{DKP-W-evolve-r})
and (\ref{DKP-W-evolve-t}) can be readily
converted to the evolution equations
\begin{gather}
  e^{\rd_r}\left(\begin{array}{cccc}
  \Psi_1 & \Psi^*_1 \\
  \Psi_2 & \Psi^*_2
  \end{array}\right)
= \left(\begin{array}{cc}
  A & B \\
  C & 0
  \end{array}\right)
  \left(\begin{array}{cccc}
  \Psi_1 & \Psi^*_1 \\
  \Psi_2 & \Psi^*_2
  \end{array}\right)
\label{DKP-lin-Psi-r}
\end{gather}
and
\begin{gather}
  \rd_{t_n}\left(\begin{array}{cccc}
  \Psi_1 & \Psi^*_1 \\
  \Psi_2 & \Psi^*_2
  \end{array}\right)
= \left(\begin{array}{cc}
  A_n & B_n \\
  C_n & D_n
  \end{array}\right)
  \left(\begin{array}{cccc}
  \Psi_1 & \Psi^*_1 \\
  \Psi_2 & \Psi^*_2
  \end{array}\right),
\label{DKP-lin-Psi-t}
\end{gather}
for the wave functions.

\subsection*{Acknowledgements}

The author is grateful to Ralf Willox and
Saburo Kakei for useful information and comments.
This work is partly supported by Grant-in-Aid for
Scientif\/ic Research No. 19540179 and No. 21540218
from the Japan Society for the Promotion of Science.

\pdfbookmark[1]{References}{ref}

\LastPageEnding

\end{document}